\documentclass[10pt,twocolumn,twoside,letterpaper,final]{IEEEtran}

\usepackage{subfigmat}

\usepackage{cite}

\ifCLASSINFOpdf
 \usepackage[pdftex]{graphicx}
\else
 \usepackage{epsfig}
\fi
\usepackage[cmex10]{amsmath}
\usepackage{amsfonts}
\usepackage{algorithmic}
\usepackage{algorithm}

\usepackage{array}

\usepackage{mdwmath}

\usepackage[caption=false,font=footnotesize]{subfig}
\usepackage{url}

\ifCLASSOPTIONdraftcls

\else

\fi

\usepackage{bm}

\usepackage{color}

\begin{document}
%
\title{Space Time MUSIC: Consistent Signal Subspace Estimation for Wide-band Sensor Arrays}
%
%
%

\author{Elio~D.~Di~Claudio, 
 Raffaele~Parisi, 
 and~Giovanni~Jacovitti 
\thanks{Manuscript received \today; revised Month, Year.}
\thanks{The authors are with the Department of Information Engineering, Electronics and Telecommunications, University of Rome ``La Sapienza'', via Eudossiana 18, 00184 Rome, Italy. (e-mail: \{elio.diclaudio, raffaele.parisi\}@uniroma1.it, gjacov@infocom.uniroma1.it)}
}%

%
%

\markboth{SUBMITTED}%
{Di Claudio \MakeLowercase{\textit{et al.}}: Space Time MUSIC: Consistent Signal Subspace Estimation for Wide-band Sensor Arrays}
%



\maketitle

\begin{abstract}
Wide-band Direction of Arrival (DOA) estimation with sensor arrays is an essential task in sonar, radar, acoustics, biomedical and multimedia applications. Many state of the art wide-band DOA estimators coherently process frequency binned array outputs by approximate Maximum Likelihood, Weighted Subspace Fitting or focusing techniques. This paper shows that bin signals obtained by filter-bank approaches do not obey the finite rank narrow-band array model, because spectral leakage and the change of the array response with frequency within the bin create \emph{ghost sources} dependent on the particular realization of the source process. Therefore, existing DOA estimators based on binning cannot claim consistency even with the perfect knowledge of the array response. In this work, a more realistic array model with a finite length of the sensor impulse responses is assumed, which still has finite rank under a space-time formulation. It is shown that signal subspaces at arbitrary frequencies can be consistently recovered under mild conditions by applying MUSIC-type (ST-MUSIC) estimators to the dominant eigenvectors of the wide-band space-time sensor cross-correlation matrix. A novel Maximum Likelihood based ST-MUSIC subspace estimate is developed in order to recover consistency. The number of sources active at each frequency are estimated by Information Theoretic Criteria. The sample ST-MUSIC subspaces can be fed to any subspace fitting DOA estimator at single or multiple frequencies. Simulations confirm that the new technique clearly outperforms binning approaches at sufficiently high signal to noise ratio, when model mismatches exceed the noise floor.
\end{abstract}


\ifCLASSOPTIONpeerreview
 \begin{center} \bfseries EDICS Category: SAM-DOAE, RAS-LCLZ. \end{center}
\fi
%
\IEEEpeerreviewmaketitle

\section{Introduction}
\label{section:intro}
\IEEEPARstart{W}{ide-band} signal processing with sensor arrays is a relevant field of research in widespread remote sensing applications. In particular, Direction of Arrival (DOA) estimation in sonar, multi-band radar, acoustic surveillance, seismics, video conferencing and Ultra-Wide Band (UWB) radio mostly involves signal bandwidths that largely violate the narrow-band assumption.

Wide-band DOA estimation is usually tackled by four main approaches:
\begin{itemize}
\item{{\it Frequency binning} \cite{SCHULTHEISS93}, performed by decomposing the wide-band signals into a set of narrow-band signals (bins), each tuned to a different frequency. The spatial information carried by a set of narrow-band \emph{spatial covariance} matrices (SCMs) is combined into a unique DOA estimate by either a multi-frequency extension of Maximum Likelihood (ML) \cite{BOHME85,SCHULTHEISS93,DORON93}, MUSIC \cite{BUCKLEY88} and Weighted Subspace Fitting (WSF) \cite{CADZOW90,VIBERG91} narrow-band DOA estimators, or a \emph{coherent focusing} stage \cite{WANG85,HUNG88,DORON92,LEE94,VALAEE95,DICLAUDIO01,YOON06}.}
\item{{\it Time Difference of Arrival} (TDOA) estimation, where wavefronts are reconstructed by fitting estimates of the differential time delays among pairs of sensors \cite{BIENATI01}.}
\item{{\it Wide-band adaptive steered beamforming}, optimized by a Minimum Variance (MV) \cite{KROLIK89} or ML criterion in the frequency domain \cite{DICLAUDIO03}.}
\item{ {\it Compressed Sensing} (CS), where the sparsity of sources is enforced by a penalty functional in a quantized DOA domain \cite{CADZOW90,WIPF07}.}
\end{itemize}

The resolving capability of TDOA approaches is limited by sampling and windowing and by the presence of multiple sources and colored noise \cite{BIENATI01}.

Wide-band steered beamforming and CS furnish very robust but inconsistent DOA estimates, respectively asymptotically biased and resolution limited.

Since parametric narrow-band DOA estimation is consistent and can cope with multi-source, correlated noise and arbitrary array geometries, there is a strong interest in extending its use to the wide-band case through frequency binning. However the observation time must be sufficiently small to comply with the assumption of stationarity of the sources and to provide an adequate number of independent snapshots for the stable SCM estimation within each bin. These contrasting requirements enforce an unavoidable spread of the binning filters in the frequency domain \cite{HARRIS78}. 

\begin{figure}
	\centering
		\includegraphics[width=3.2in]{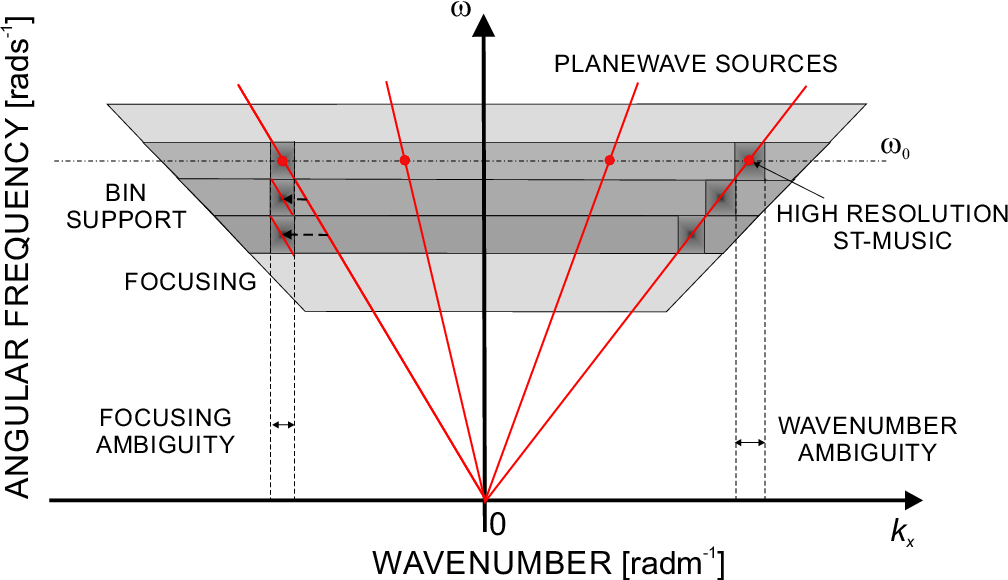}
	\caption{The spectral support of four far field noiseless and uncorrelated sources impinging on an ULA in the half plane whose coordinates are the the wavenumber $k_x$ and the angular frequency $\omega > 0$ \protect \cite{CAPON69}. The wavenumber (and therefore the DOA) uncertainties due to the finite bin width and to a typical focusing process \protect \cite{HUNG88} are illustrated.}
	\label{fig:Fig1}
\end{figure}

The problem is sketched in Fig. \ref{fig:Fig1} for a uniform linear array (ULA). The spectral support of each noiseless plane wave source propagating at speed $u$ and impinging from azimuth $\theta$, referred to broadside, is a Dirac wall based on the line ${k_x} = - {{\omega \sin \left( \theta \right)} \mathord{\left/ {\vphantom {{\omega \sin \left( \theta \right)} u}} \right. \kern-\nulldelimiterspace} u}$ in the plane formed by the wavenumber $k_x$ and the angular frequency $\omega$, whose mass is modulated by the source spectrum \cite{CAPON69}. 

In binning approaches, best alignments of DOA estimates from different bins are inferred \cite{CADZOW90,HUNG88}. However, the DOA ambiguity caused by variations of the source spectrum and the steering vector \cite{SCHMIDT86} within each bin and the \emph{spectral leakage} among adjacent bins due to window side-lobes cannot be recovered, since the narrowband array model does not account for them. In particular:
\begin{itemize}
\item {Spatially spread, uncalibrated {\em ghost sources} emerge from noise background for increasing signal to noise ratio (SNR) and/or sample size \cite{DICLAUDIO01,DICLAUDIO05,DICLAUDIO13}.}
\item{Therefore, the \emph{finite rank} structure of the narrow-band array model \cite{SCHMIDT86} does not hold within each bin.}
\item{The model becomes unidentifiable and the detection of the source number \cite{WAX85} is an ill-posed problem.}
\item{The model mismatch generates a \emph{signal-dependent} DOA bias.}
\item{For random sources, an excess DOA variance appears at high SNR \cite{DICLAUDIO01} and may even hamper the convergence of optimal DOA estimators \cite{SWINDLEHURST93}, such as ML \cite{DORON93} and WSF \cite{CADZOW90,STOICA90a,VIBERG91}.}
\end{itemize}

This mismatch undermines the \emph{consistency} of the existing narrow-band DOA estimators \emph{even in the absence of noise} used on frequency bins and is especially relevant for wide spatial apertures or when wide-band, short-time acquisitions are required. It adds up to calibration and focusing mismatches, all relevant at \emph{high} SNR \cite{SWINDLEHURST92,SWINDLEHURST93}.

Consistency guarantees that the limit DOA solution for vanishing noise or increasing the sample size is unique and satisfies the underlying physical model, regardless of signal realizations. This is important for scientific experiments, array calibration, target tracking and nulling static interference in telecommunications. 

Our goal is to obtain a \emph{consistent estimate of the signal subspaces, asymptotically free from artifacts} to replace the SCM counterparts. With reference to Fig. \ref{fig:Fig1}, it is clear that this task requires a joint parametric modeling in space and time. 

In \cite{KROLIK90} high resolution, narrow-band SCMs were estimated by suppressing the spectral leakage by a bank of MV, distortion-less, delay and sum (DS) beamformers, tuned to the same frequency. This approach involves cross-temporal moments through the \emph{space-time array covariance matrix} (STCM), which only requires time sampling of sensor outputs at a sufficient rate. The STCM exhibits fast statistical convergence and low bias even for short data records. However, the resulting Capon SCM estimator of \cite{KROLIK90} suffers from the small sample issues of MV beamforming \cite{LI05} and its structure was not analyzed for DOA estimation purposes.

In this work, starting from the finite rank property of the wide-band source signature on the STCM, under the assumption of finite lengths of the source to sensor impulse responses \cite{KROLIK90}, an extension of the narrow-band DOA identifiability conditions \cite{SCHMIDT86,DORON93} shows that consistent wide-band DOA estimation is theoretically feasible by space time ML or WSF approaches. However, such estimators require the knowledge of the number and the impulse responses of all sources and a very accurate coarse DOA estimation, which is at least un-practical. 

In \cite{DICLAUDIO13}, a MUSIC \cite{SCHMIDT86} type condition allowed to asymptotically recover \emph{finite rank} narrow-band signal subspaces at \emph{arbitrary} frequencies from the the \emph{wide-band dominant eigenvectors} of the STCM (or any other statistic sharing a similar structure \cite{GELLI03,GARDNER06}), rather than from filtered array data. Earlier versions of this generalized \emph{Space-Time MUSIC} (ST-MUSIC) \cite{DICLAUDIO13,DICLAUDIO14} subspace estimator did not allow for a sound statistical estimation of the number of sources and had a slowly decaying asymptotic bias, which hampered the numerical stability of the DOA estimates, especially at extreme SNR and with small samples, without the use of heuristic regularization \cite{LEDOIT04}. 

Therefore, a novel ST-MUSIC subspace estimator, sharing similarities with the Signal Subspace-MUSIC (SS-MUSIC) \cite{MCCLOUD02}, was derived from the ML formulation of an \emph{inversion} problem. The ML setting optimizes the statistical efficiency and increases the robustness of the sample signal subspace at extreme SNR, thus allowing the development of subspace rank estimators, based on Akaike \cite{AKAIKE74} and Bayesian \cite{WAX85,BURNHAM04} Information Theoretic Criteria (AIC and BIC respectively).

A perturbative analysis of the ML ST-MUSIC signal subspace estimate established its consistency and the optimal weighting in the WSF sense \cite{VIBERG91}. Therefore, the sample ML ST-MUSIC subspaces at different frequencies can \emph{replace} the inconsistent SCM counterparts in any narrow-band or wide-band subspace based DOA estimator \cite{CADZOW90,VALAEE95,DICLAUDIO01,YOON06} regardless of the spatial coherency of sources.

Numerical simulations of the ML ST-MUSIC subspaces with WSF based DOA estimation schemes \cite{VIBERG91,DICLAUDIO01} show a definite statistical performance improvement at high SNR of the ML ST-MUSIC subspaces over the classical counterparts, approaching the Cramer Rao DOA variance bound, at the cost of a slightly increased low SNR threshold in some cases, whose causes are discussed, and higher computation burden.

After a notation Sect. \ref{section:Notation}, the finite length wide-band array response model is analyzed in Sect. \ref{section:arraymodel} for DOA estimation purposes. In Sect. \ref{section:SpaceTimeMUSIC} the ST-MUSIC concepts are developed. In Sect. \ref{sec:STMUSICEstimator}, the ML ST-MUSIC subspace estimator is developed and AIC and BIC rank estimators are deduced, based on the first order finite sample perturbation models of the STCM. In Sect. \ref{sec:ComputationalAnalysis} the computational analysis is sketched. In Sect. \ref{sec:results} the validity of the ML ST-MUSIC model is assessed by numerical simulations. Finally, conclusion is drawn in Sect. \ref{sec:conc}.

\section{Notation}\label{section:Notation}
Throughout the paper matrices are indicated by boldface, capital letters, vectors by boldface, lowercase letters. The transpose of matrix $\bf A$ is indicated by ${\bf A}^T$, the conjugate by ${{\bf{A}}^*}$, the Hermitian transpose by ${\bf A}^H$. The pseudo-inverse of $\bf A$ is $\bf A^\dagger$. ${\bf I}_M$ is the square identity matrix of size $M$. The operator ${\rm diag}\{{\bf A}\}$ creates a column vector with the main diagonal of $\bf A$, ${\rm diag}\{{\bf a}\}$ creates a diagonal matrix with the elements of vector ${\bf a}$ placed on its main diagonal. $\mathop{\rm {trace}}({\bf A})$ is the trace of matrix $\bf A$. $\mathop{\rm {det}}({\bf A})$ is the determinant of $\bf A$. ${\bf A}\otimes{\bf B}$ is the Kronecker product between matrices $\bf A$ and $\bf B$. $\delta_{kl}$ is the Kronecker symbol, equal to one if $k=l$ and zero elsewhere. $j=\sqrt{-1}$ is the imaginary unit. Sub-matrices are indexed by MATLAB-like conventions \cite{GOLUB89}. For instance, ${\bf A}(:,k)$ is the $k-$th column of matrix $\bf A$ and ${\bf A}(1:m,1:n)$ is the upper left submatrix of $\bf A$ of size $m\times n$. The operator $\ast$ indicates the temporal convolution of two signals. ${\mathop{\rm E}\nolimits} \left\{ x \right\}$ is the expected value of the random variable $x$ and ${\mathop{\rm var}\nolimits} \left( {x} \right)$ denotes its variance. The covariance between the random variables $x$ and $y$ is indicated by ${\mathop{\rm cov}\nolimits} \left( {x,y} \right)$ . The Frobenius norm \cite{GOLUB89} of $\bf A$ is indicated by ${\left| {\bf{A}} \right|_F}$. Sample quantities are denoted by a \emph{hat} superscript (e.g., ${\hat{\bf A}}$).

%

\section{Array model}
\label{section:arraymodel}
A sensor array with $M$ sensors is immersed in a wave-field and receives the signals $s_d(t)$ ($d=1,2,\hdots,D$), radiated by $D\!<\!M$ wide-band point sources, whose DOAs are characterized by the generic coordinate vectors $\bm{\theta}_d$. Provided that the sensors and the propagating medium are linear, the signals $x_m(t)$ ($m=1,2,\hdots,M$), received by the $M$ sensors at time $t$, are modeled as 
\begin{equation}
{x_m}\left( t \right) = \sum\limits_{d = 1}^D {{h_m}} \left( {t;{\bm\theta _d}} \right) \ast {s_d}\left( t \right) + {v_m}\left( t \right)
\label{eqn:wbsensorsignal}
\end{equation}
where $h_m\left(t;\bm{\theta}_d\right)$ is the impulse response from the $d$-th source to the $m$-th sensor, including the propagation channel (i.e., the Green's function) and the front end filtering\footnote{Following this formulation, $h_m\left(t;\theta_d\right)$ may collect the effects of \emph{multiple coherent reflections} originated by the same driving signal $s_d\left(t\right)$ \cite{GINGRAS93}.}, and $v_m\left(t\right)$ is the additive sensor noise term, assumed as statistically independent of source signals. All signal components can be either complex envelopes with respect to an angular frequency $\omega_0$, or real valued base-band signals. For a down conversion carrier angular frequency $\omega_0$ and a sampling period $T$, the array response at the discrete time angular frequency $\nu$ coincides with the response to a continuous time angular frequency $\omega\left({\nu}\right)=\nu/T+\omega_0$. 

The sequences of consecutive $P$ array output samples ${x_m}\left( n \right) = {\left. {{x_m}\left( t \right)} \right|_{t = nT}}$ for $n=1,2,\ldots,N$ are collected in a \emph{space time snapshot} (STS) of dimension $MP\times 1$
\begin{equation}
{{\bf{x}}_{ST}}\left( n \right) = {\left[ {\begin{array}{*{20}{c}}
{{\bf{x}}_1^T\left( n \right)}&{{\bf{x}}_2^T\left( n \right)}& \cdots &{{\bf{x}}_M^T\left( n \right)}
\end{array}} \right]^T}
\label{eqn:spacetimesnapshot}
\end{equation}
where
\begin{equation}
{{\bf{x}}_m}\left( n \right) = {\left[ {\begin{array}{*{20}{c}}
{{x_m}\left( n \right)}&{{x_m}\left( {n - 1} \right)}& \cdots &{{x_m}\left( {n - P + 1} \right)} 
\end{array}} \right]^T}\;.
\label{eqn:delayedinputs}
\end{equation}

Each discrete time sensor impulse response ${\bf h}_m\left(\bm{\theta}_d\right)$ ($m=1,2,\hdots,M$, $d=1,2,\hdots,D$) represents the discrete time counterpart of $h_m\left(t;\bm{\theta}_d\right)$ and is described by the row vector
\begin{equation}
\begin{array}{c}
{{\bf{h}}_m}\left( {{{\bm{\theta }}_d}} \right) = 
\left[ {\begin{array}{*{20}{c}}
{{h_{m,d}}\left( 0 \right)}&{{h_{m,d}}\left( 1 \right)}& \cdots &{{h_{m,d}}\left( {{L_{m,d}} - {1}} \right)} 
\end{array}} \right]\;.
\end{array}
\label{eqn:impulseresponse}
\end{equation}

Its duration $L_{m,d}$ may be infinite, if ${{\bf{h}}_m}\left( {{{\bm{\theta }}_d}} \right)$ contains poles \cite{CICCHETTI16}. However, $\left| {{h_{m,d}}\left( n \right)} \right|$ typically quickly decays with $n$ well under the noise floor and it makes sense to assume a common \emph{effective overall length} $L_d=\max_{1\le m\le M}L_{m,d}$ for all ${\bf h}_m\left(\bm{\theta}_d\right)$ related to the generic $d$-th source.

Each impulse response ${\bf h}_m\left(\bm{\theta}_d\right)$, zero padded to $L_d$, is used to build a Toeplitz \emph{convolution matrix} of size $P\times (P+L_d-1)$
\begin{equation}
{\bf H}_m\left(\bm{\theta}_d\right)=\left[
\begin{matrix}
{\bf h}_m\left(\bm{\theta}_d\right) & 0 & \cdots & 0 \cr
0 & {\bf h}_m\left(\bm{\theta}_d\right) & \cdots & 0 \cr
\cdots & \cdots & \cdots & \cdots \cr
0 & 0 & \cdots & {\bf h}_m\left(\bm{\theta}_d\right) \cr
\end{matrix}\right]
\label{eqn:impulseresponsematrix}
\end{equation}
for $m = 1,2,\ldots,M$. Finally, the STS \eqref{eqn:spacetimesnapshot} is compactly expressed as
\begin{equation}
{\bf x}_{ST}\left({n}\right)=\sum_{d=1}^{D}{\bf H}\left(\bm{\theta}_d\right){\bf s}_d\left({n}\right)+{\bf v}\left({n}\right)
\label{eqn:spacetimedatamatrix}
\end{equation}
where ${\bf H}\left(\bm{\theta}_d\right)$ stacks ${\bf H}_m\left(\bm{\theta}_d\right)$ for $m=1,\cdots,M$
and is the multi-channel convolution matrix of the sampled $d$-th source signal
\begin{equation}
{{\bf{s}}_d}\left( n \right) = {\left[ {\begin{array}{*{20}{c}}
{{{\bf{s}}_d}\left( {n} \right)}&{{{\bf{s}}_d}\left( {n-1} \right)}& \cdots &{{{\bf{s}}_d}\left( {n - P - L_d + 2} \right)}
\end{array}} \right]^T}
\label{eqn:delayedsignalvector}
\end{equation}
and the $MP \times 1$ vector
\begin{equation}
{\bf{v}}\left( n \right) = {\left[ {\begin{array}{*{20}{c}}
{{\bf{v}}_1^T\left( n \right)}&{{\bf{v}}_2^T\left( n \right)}& \cdots &{{\bf{v}}_M^T\left( n \right)}
\end{array}} \right]^T}\label{eqn:noisesnapshot}
\end{equation}
collects the additive sensor noise samples
\begin{equation}
{{\bf{v}}_m}\left( n \right) = {\left[ {\begin{array}{*{20}{c}}
{{v_m}\left( {n} \right)}&{{v_m}\left( {n-1} \right)}& \cdots &{{v_m}\left( {n- P + 1} \right)} 
\end{array}} \right]^T}\;.
\label{eqn:delayednoisevector}
\end{equation}

Equations \eqref{eqn:impulseresponsematrix} and \eqref{eqn:spacetimedatamatrix} show that the rank of ${\bf H}\left({\bm{\theta}_d}\right)$ is upper bounded by $P+L_d-1$ \cite{KROLIK90}. The SCM model \cite{SCHMIDT86} can be recovered by assuming $L_d=1$ and setting $P=1$.

\subsection{Space time covariance matrix }
\label{section:STCM model}
To tackle the consistency problem, cross-temporal moments are required to model the spectral spread. To this purpose, the \emph{Space-Time Covariance Matrix} (STCM) ${\bf R}_{ST}=\mathop{E}\left\{ {{\bf x}_{ST}\left({n}\right){\bf x}^H_{ST}\left({n}\right)}\right\}$ \cite{KROLIK90} is a natural starting point for exploiting the model \eqref{eqn:spacetimedatamatrix}, because it collects all available second-order information for zero mean, circularly complex signals and noise with bounded fourth-order moments, if $P > \mathop {\max }\limits_d \left( {{L_d} + {L_{s,d}}} \right)$, where ${L_{s,d}}$ is the correlation length of the $d$th source\footnote{The use of other statistics based on \eqref{eqn:spacetimedatamatrix}, such as biased \cite{DICLAUDIO13}, structured \cite{DICLAUDIO11} or cyclic correlation matrices \cite{GELLI03}, is deferred to future work.}. It is worth to point out that in the same scenario the SCM estimate would require the use of a DFT of length $P_w \gg P > \mathop {\max }\limits_d \left( {{L_d} + {L_{s,d}}} \right)$, because its bias slowly decays as $P_w^{-1}$ \cite{HARRIS78,HUNG88,SCHULTHEISS93}. 

In the sequel, the vector of source signals 
\begin{equation}
{\bf{s}}\left( n \right) = {\left[ {\begin{array}{*{20}{c}}
{{\bf{s}}_1^T\left( n \right)}& \cdots &{{\bf{s}}_D^T\left( n \right)}
\end{array}} \right]^T}
\label{eqn:signalsnapshot}
\end{equation}
is assumed as a realization of a circular, wide sense stationary process with zero mean and covariance
\begin{equation}
{\bf S}=\mathop{E} \left\{ {{\bf s}\left({n}\right){\bf s}^H\left({n}\right)}\right\} = \left[
\begin{matrix}
{\bf S}_{11} & \cdots & {\bf S}_{1D} \cr
\cdots & \cdots & \cdots \cr
{\bf S}_{1D}^H & \cdots & {\bf S}_{DD} \cr
\end{matrix}\right]
\label{eqn:signalcovariancematrix}
\end{equation}
whose blocks ${\bf S}_{kl}=\mathop{E}\left\{ {{\bf s}_k\left({n}\right){\bf s}_l^H\left({n}\right)}\right\}$ are the cross-covariance matrices between each pair of impinging signals. 

Sensor noise is assumed zero mean, circular and independent of source signals, with STCM $\mathop{E} \left\{ {{\bf v}\left({n}\right){\bf v}^H\left({n}\right)} \right\} = \lambda_v {\bf R}_{vv}$ where $\lambda_v \ge 0$. Then, the \emph{array STCM} is 
\begin{equation}
{\bf R}_{ST}=\mathop{E}\left\{ {{\bf x}_{ST}\left({n}\right){\bf x}^H_{ST}\left({n}\right)}\right\} = {\bf H}_{ST}{\bf S}{\bf H}^H_{ST}+\lambda_v{\bf R}_{vv}
\label{eqn:arraySTCM}
\end{equation}
where ${{\bf{H}}_{ST}} = \left[ {\begin{array}{*{20}{c}}
{{\bf{H}}\left( {{{\bm{\theta }}_1}} \right)}& \cdots &{{\bf{H}}\left( {{{\bm{\theta }}_D}} \right)}
\end{array}} \right]$.

\subsection{Space time signal and noise subspaces}
\label{sec:Space time signal and noise subspaces}
For known ${\bf R}_{vv} = {\bf R}_v{\bf R}_v^H$, the orthonormal and complementary bases ${\bf E}_s$ for the \emph{wide-band signal subspace} and ${\bf E}_v$ for the \emph{wide-band noise subspace} are classically defined by the eigen-decomposition (EVD) \cite{GOLUB89} of the whitened STCM
\begin{equation}
{\bf R}_v^{-1}{\bf R}_{ST}{\bf R}_v^{-H} = {\bf E}_s{\bf \Lambda}_{s}{\bf E}_s^H+\lambda_v{\bf E}_v^{}{\bf E}_v^H
\label{eqn:GEVD}
\end{equation}
where the diagonal matrix ${\bf \Lambda}_{s} = \mathop{\rm{diag}} \{\lambda_1,\hdots,\lambda_{\eta}\}$ contains the dominant $\eta$ eigenvalues in non-increasing order (i.e., $\lambda_1\ge\lambda_2\ge\hdots\ge\lambda_{\eta} > \lambda_v$). In particular, the subspace ${\bf E}_s$ has dimension
\begin{equation}
\eta \le \sum_{d=1}^D \left({P+L_d-1}\right) \le D\left[P+{\max_d \left({L_d} \right)-1}\right] 
\label{eqn:WSSdimension}
\end{equation}
and the column span of ${\bf E}_s$ lies within the column span of ${\bf H}_{ST}$, so that \cite{VIBERG91}
\begin{equation}
{{\bf{E}}_s} = {\bf{R}}_v^{ - 1}\left[ {\begin{array}{*{20}{c}}
{{\bf{H}}\left( {{{\bm \theta }_1}} \right)}& \cdots &{{\bf{H}}\left( {{{\bm \theta}_D}} \right)}
\end{array}} \right]\left[ {\begin{array}{*{20}{c}}
{{{\bf{C}}_1}}\\
 \vdots \\
{{{\bf{C}}_D}}
\end{array}} \right]
\label{eqn:columnspaceofEs}
\end{equation}
where the blocks ${\bf C}_d$ for $d = 1,2,\ldots,D$ have size $\left(P+L_d-1\right)\times\eta$ and point out the contributes of each source.

\subsection{DOA identifiability from the STCM by ML and WSF approaches}
\label{subsec:DOAIdentifiabilityFromTheSTCMbyMLandWSFapproaches}
Model \eqref{eqn:arraySTCM} is theoretically amenable to the straight extension of ML \cite{BOHME85,SCHULTHEISS93,DORON93} and WSF \cite{CADZOW90,VIBERG91} narrow-band DOA estimators\footnote{For zero mean circular Gaussian signals and noise the STCM is a sufficient statistic for DOA and $\bf S$ parameters of model \eqref{eqn:spacetimedatamatrix}.}, assuming for instance the \emph{exact} knowledge of the matrix impulse response ${\bf H}(\bm{\theta})$ for all $\bm{\theta}$, of the noise covariance ${\bf R}_{vv}$ up to a positive scalar, and of the exact number of paths to search for. In addition, the source combination which achieves the target ${\bf R}_{ST}$ must be unique \cite{SCHMIDT86}. 

In particular, the rank $\eta$ of ${\bf E}_s$ must remain smaller than $MP$ for the existence of ${\bf E}_v$, crucial for the unambiguous identification of $\lambda_v$ and DOAs \cite{SCHMIDT86,VIBERG91}. By \eqref{eqn:spacetimedatamatrix}, each uncorrelated source contributes to $\eta$ by a variable quantity ranging from one (e.g., for a pure complex sinusoid) \cite{DICLAUDIO13} up to $P+L_d-1$ for a random source spanning the full array bandwidth. Since for $D$ full-band sources $\eta>DP$, the number of resolvable wide-band sources is \emph{strictly smaller} than $M$. Further limitations on the identifiability may be imposed by array ambiguities and spatial coherence \cite{SCHMIDT86}. 

On the contrary, narrow-band sources generate only a few significant eigenvalues and proper techniques \cite{SU83,WAX84,MOHAN08} can already estimate more DOAs than sensors if the source spectra do not overlap. 

However, the wide-band ML and WSF approaches based on \eqref{eqn:spacetimedatamatrix} lead to awkward, huge-dimensional problems that need to be initialized by accurate (within half beam-width or less \cite{BOHME85}) prior DOA estimates for successful local convergence. In particular, the difficulty of measuring the full ${\bf H}(\bm{\theta})$ \cite{CICCHETTI16} motivated us to develop approaches based only on the knowledge of the array harmonic response at a discrete set of temporal frequencies. 

\section{Space Time MUSIC}
\label{section:SpaceTimeMUSIC}
In \cite{DICLAUDIO13}, a set of \emph{consistent} and optimally weighted in the WSF sense \cite{VIBERG91} \emph{narrow-band signal subspace bases} at \emph{arbitrary} frequencies were obtained directly from ${{\bf{E}}_s}$ through an extension of the MUSIC paradigm. This solution, referred to as \emph{Space-Time MUSIC} (ST-MUSIC), bypasses traditional filtering, SCM building and EVD stages. In the sequel, the STCM signal model at frequency $\nu$ is generalized to the non spherical noise case and the ML ST-MUSIC subspace estimate is derived from the solution of an inversion problem from ${{\bf{E}}_s}$.

The space-time array response to a pure unit amplitude, complex sinusoid $s\left({n}\right) = e^{j\nu n}$, impinging from the generic angle $\bm \theta$, can be written as
\begin{equation}
	{{{\bf{x}}_{ST}}\left( n \right) = {{\bf{a}}_{ST}}\left( {\nu ,{\bm{\theta }} } \right){e^{j\nu n}}}
	\label{eqn:sinusoidalresponse}
\end{equation}
where ${\bf a}_{ST}\left({\nu,\bm{\theta}}\right)$, referred to as the \emph{space time steering vector} (STSV), has size $MP \times 1$ and is defined by the Kronecker product 
\begin{equation}
{{\bf{a}}_{ST}}\left( {\nu ,{\bm{\theta }}} \right) = {\bf{a}}\left( {\nu ,{\bm{\theta }}} \right) \otimes {{\bf{e}}_P}\left( \nu \right) = {P^{1/2}}{\bf{E}}\left( \nu \right){\bf{a}}\left( {\nu ,{\bm{\theta }}} \right)
\label{eqn:spacetimearrayresponse}
\end{equation}
where ${\bf a}\left(\nu,\bm{\theta}\right)$ is the classical $M \times 1$ narrow-band array steering vector at frequency $\nu$ \cite{SCHMIDT86} and ${{\bf{e}}_P}\left( {\nu} \right) = {\left[ {\begin{array}{*{20}{c}}
1&{{e^{ - j\nu }}}& \cdots &{{e^{ - j\left( {P - 1} \right)\nu }}}
\end{array}} \right]^T}$ is a vector of size $P \times 1$ \cite{KROLIK90,DICLAUDIO14}. The last equality in \eqref{eqn:spacetimearrayresponse} defines the $MP\times M$ orthogonal matrix
\begin{equation}
{\bf E}\left({\nu}\right)=P^{-1/2}{\bf I}_M\otimes{\bf e}_P\left({\nu}\right)
\label{eqn:frequencysubspace}
\end{equation}
as a basis for the subspace spanned by any ${\bf a}_{ST}\left({\nu,\bm{\theta}}\right)$.

To put into evidence the relationship between ${{\bf{a}}_{ST}}\left( {\nu ,{\bm{\theta }}} \right)$ and ${{\bf{H}}}\left( {{\bm{\theta }}} \right)$, let us define the unitary matrix of size ${P+L_d-1}$
\begin{equation}
	{{\bf{D}}}(\nu ) = \left[ {\begin{array}{*{20}{c}}
\left({P+L_d-1}\right)^{-1/2}{{{\bf{e}}_{P+L_d-1}}\left( \nu \right)}&{{{\bf{D}}_{\bot }}\left( \nu \right)}
\end{array}} \right]
\label{eqn:unitaryD}
\end{equation}
where ${{\bf{D}}_{\bot }}\left( \nu \right)$ is the orthogonal complement \cite{GOLUB89} to ${{{\bf{e}}_{P+L_d-1}}\left( \nu \right)}$. 

After posing ${{\bf{s}}_{F,d}}\left( {n,\nu } \right) = {\bf{D}}_{{L_d} + P - 1}^{ H}\left( \nu \right){{\bf{s}}_d}\left( n \right)$ and 
\begin{equation}
\label{eqn:modulatedresponse}
\begin{array}{*{20}{c}}
{{\bf{H}}\left( {\nu ,{{\bm{\theta }}_d}} \right) = {\bf{H}}\left( {{{\bm{\theta }}_d}} \right){{\bf{D}}_{{L_d} + P - 1}}\left( \nu \right) = }\\
{\left[ {\begin{array}{*{20}{c}}
{{{\bf{a}}_{ST}}\left( {\nu ,{{\bm{\theta }}_d}} \right)}&{{\bf{H}}\left( {\nu ,{{\bm{\theta }}_d}} \right){{\bf{D}}_ \bot }\left( \nu \right)}
\end{array}} \right]}
\end{array}
\end{equation}
the response \eqref{eqn:spacetimedatamatrix} of the $d$-th source signal after noise whitening can be rewritten by \eqref{eqn:sinusoidalresponse} as
\begin{equation}
	\begin{array}{c}
{\bf{R}}_v^{ - 1}{{\bf{x}}_{ST}}\left( n \right) = {\bf{R}}_v^{ - 1}{\bf{H}}\left( {\nu ,{{\bm{\theta }}_d}} \right){{\bf{s}}_{F,d}}\left( {n,\nu } \right)\; = \\
{\bf{R}}_v^{ - 1}\left[ {\begin{array}{*{20}{c}}
{{P^{{1 \mathord{\left/
 {\vphantom {1 2}} \right.
 \kern-\nulldelimiterspace} 2}}}{\bf{E}}\left( \nu \right){\bf{a}}\left( {\nu ,{{\bm{\theta }}_d}} \right)}&{{\bf{H}}\left( {\bm{\theta} _d} \right){{\bf{D}}_ \bot }\left( \nu \right)}
\end{array}} \right]{{\bf{s}}_{F,d}}\left( {n,\nu } \right) \;.
\end{array}
\label{eqn:noiselessspacetimeresponse}
\end{equation}
 
To describe the effects of noise whitening on the signal model at frequency $\nu$, let us define the following full QR decomposition (QRD) 
\begin{equation}
{P^{1/2}}{\bf{R}}_v^{ - 1}{\bf{E}}\left( \nu  \right) = \left[ {\begin{array}{*{20}{c}}
{{\bf{A}}\left( \nu  \right)}&{{{\bf{A}}_ \bot }\left( \nu  \right)}
\end{array}} \right]\left[ {\begin{array}{*{20}{c}}
{{\bf{R}}_v^{ - 1}\left( \nu  \right)}\\
{\bf{0}}
\end{array}} \right]
\label{eqn:whitenedresponse}
\end{equation}
where ${\bf A}\left({\nu}\right)$ is the $M$ dimensional \emph{frequency subspace basis} at $\nu$, ${\bf A}_{\bot} \left({\nu}\right)$ is its orthogonal complement and ${\bf{R}}_v^{ - 1} \left( {\nu} \right)$ is a square upper triangular matrix of size $M$, which defines the following \emph{noise whitened narrow-band steering vector} of size $M \times 1$
\begin{equation}
	{\bf b}\left({\nu,\bm{\theta}}\right) = {\bf{R}}_v^{ - 1} \left( {\nu} \right) {\bf a}\left({\nu,\bm{\theta}}\right) \;.
	\label{eqn:whitenedsteeringvector}
\end{equation}

Combining \eqref{eqn:noiselessspacetimeresponse}, \eqref{eqn:whitenedresponse} and \eqref{eqn:whitenedsteeringvector} leads to the decomposition
\begin{equation}
\begin{array}{c}
\left[ {\begin{array}{*{20}{c}}
{\begin{array}{*{20}{c}}
{{{\bf{A}}^H}\left( \nu \right)}\\
{{\bf{A}}_ \bot ^H\left( \nu \right)}
\end{array}}
\end{array}} \right]{\bf{R}}_v^{ - 1}{\bf{H}}\left( {\nu ,{{\bm{\theta }}_d}} \right){{\bf{s}}_{F,d}}\left( {\nu ,n} \right) = \\
\left[ {\begin{array}{*{20}{c}}
{{\bf{b}}\left( {\nu ,{{\bm{\theta }}_d}} \right)}&{{{\bf{B}}_{12}}\left( {\nu ,{{\bm{\theta }}_d}} \right)}\\
{\bf{0}}&{{{\bf{B}}_{22}}\left( {\nu ,{{\bm{\theta }}_d}} \right)}
\end{array}} \right]{{\bf{s}}_{F,d}}\left( {\nu ,n} \right)\;.
\end{array}
\label{eqn:generalSTbin}
\end{equation}

The first $M$ rows contain the sum of the \emph{desired}, rank one signal component at frequency $\nu$, characterized by ${\bf b}\left({\nu,\bm{\theta}_d}\right)$, and of a \emph{spectral leakage component} transmitted by the $M\times\left({P+L_d-2}\right)$ matrix ${\bf B}_{12}\left({\nu,\bm{\theta}_d}\right)$. The latter term arises from the finite array aperture and represents the multiple rank response of a \emph{spatially spread}, uncalibrated ghost source \cite{DICLAUDIO01}. Its strength depends on the spectrum of $s_d \left( {n} \right)$ and its spatial spread on the change of ${\bf a}\left(\nu,\bm{\theta}\right)$ with frequency, which increases with the fractional bandwidth and the array aperture.

Finally, the isolated spectral leakage components can be observed in the subspace ${\bf A}_{\bot}\left({\nu}\right)$ through the transfer matrix ${\bf B}_{22}\left({\nu,\bm{\theta}_d}\right)$.

Eq. \eqref{eqn:generalSTbin} is immediately generalized to the entire ${\bf E}_s$ \eqref{eqn:columnspaceofEs} as
\begin{equation}
\left[
\begin{matrix}
{\bf A}^H\left({\nu}\right)\cr
{\bf A}_{\bot}^H\left({\nu}\right)\cr
\end{matrix}
\right]
{\bf E}_s=
\left[
\begin{matrix}
{{\bf{B}}_{11}}\left( {\nu } \right) & {\bf B}_{12}\left({\nu}\right)\cr
{\bf 0} & {\bf B}_{22}\left({\nu}\right)\cr
\end{matrix}
\right]
\left[
\begin{matrix}
{\bf C}_{F1}\left({\nu}\right)\cr
{\bf C}_{F2}\left({\nu}\right)\cr
\end{matrix}
\right]
\label{eqn:generalWSS}
\end{equation}
where ${{\bf{B}}_{11}}\left( {\nu } \right) = \left[ {\begin{array}{*{20}{c}}
{{\bf{b}}\left( {\nu ,{{\bm{\theta }}_1}} \right)}& \cdots &{{\bf{b}}\left( {\nu ,{{\bm{\theta }}_D}} \right)}
\end{array}} \right]$ is the $M\times D\left( {\nu } \right)$ narrow-band array transfer matrix at frequency $\nu$ of the $D\left( {\nu } \right) \le D$ sources with \emph{non-zero spectra} at the same $\nu$, combined by the matrix ${\bf C}_{F1}\left({\nu}\right)$ of size $D\left( {\nu } \right)\times \eta$. 

Finally the mixing matrix ${\bf C}_{F2}\left({\nu}\right)$ of size $\sum_{d=1}^D\left({P+L_d-2}\right)\;\times \eta$ models the spectral leakage observable through the transfer matrices
\begin{equation}
	\begin{array}{c}
{{\bf{B}}_{12}}\left( \nu  \right) = \left[ {\begin{array}{*{20}{c}}
{{{\bf{B}}_{12}}\left( {\nu ,{{\bm{\theta }}_1}} \right)}& \cdots &{{{\bf{B}}_{12}}\left( {\nu ,{{\bm{\theta }}_D}} \right)}
\end{array}} \right]\\
{{\bf{B}}_{22}}\left( \nu  \right) = \left[ {\begin{array}{*{20}{c}}
{{{\bf{B}}_{22}}\left( {\nu ,{{\bm{\theta }}_1}} \right)}& \cdots &{{{\bf{B}}_{22}}\left( {\nu ,{{\bm{\theta }}_D}} \right)}
\end{array}} \right]
\end{array}
\label{eqn:B11AndB22}
\end{equation}
that mark the departure of the convolutional signal model \eqref{eqn:generalWSS} from the ideal narrow-band one \cite{SCHMIDT86}.

\subsection{ST-MUSIC as an Inversion Problem}
\label{sec:MUSICAsAnInversionProblem}
With reference to \eqref{eqn:generalWSS}, a consistent narrow-band signal subspace estimator must asymptotically recover for $N/MP \to \infty $ an $M \times \kappa \left( \nu \right) $ linearly independent basis ${{\bf{B}}_s}\left( \nu \right)$ (with $ \kappa \left( \nu \right)\le D \left( \nu \right) < M$) within the span of ${{\bf{B}}_{11}}\left( {\nu } \right)$ and a full rank, square \emph{subspace weighting matrix} ${\bf{C}_s}\left( \nu \right)$ of size $\kappa \left( \nu \right)$, chosen to optimize the statistical accuracy, satisfying the WSF equation 
\begin{equation}
	{{\bf{B}}_s}\left( \nu \right){\bf{C}}_s \left( \nu \right) = {{\bf{B}}_{11}}\left( \nu \right){{\bf C}_{11}}\left( \nu \right)
	\label{eqn:narrowbandWSF}
\end{equation}
for a proper full rank mixing matrix ${{\bf C}_{11}}\left( \nu \right)$ of size $D \left( \nu \right) \times \kappa \left( \nu \right)$ \cite{VIBERG91}. 

In this formulation $\kappa \left( {\nu } \right)$ is the number of uncorrelated source signals that are \emph{active} at $\nu$ \cite{MOHAN08}. As for the SCM case, \emph{spatially coherent} (multipath) sources with $L_d+L_{s,d} < P$ have a \emph{single rank} signature in ${{\bf{B}}_s}\left( \nu \right)$ and the DOA identifiability is subjected to the same limitations \cite{VIBERG91}. In particular, the steering vectors of the fully coherent arrivals must be calibrated and ${\bf{B}}_s \left( \nu \right)$ must be analyzed by a multi-dimensional (e.g., WSF type) DOA estimator \cite{STOICA90a,VIBERG91} or subjected to spatial or frequency smoothing \cite{SHAN85,HUNG88,DICLAUDIO01}.
 
As usual, for unambiguous DOA identification, it is required that $D\left( {\nu } \right) < M$ and that any subset of $D\left( {\nu } \right)$ steering vectors ${\bf b}\left({\nu,\bm{\theta}}\right)$ is linearly independent, at least in a neighborhood of the source DOAs \cite{SCHMIDT86}. 

A viable method for consistently recovering ${\bf B}_s \left( {\nu } \right)$ from \eqref{eqn:generalWSS} is to apply a $\eta \times \kappa \left({\nu } \right)$ \emph{inversion weight matrix} $\bf{W} \left( {\nu } \right) $ to the right side of ${{\bf{E}}_s}$ to find a $M$ dimensional basis entirely lying within the span of $\bf{A}\left({\nu}\right)$:
\begin{equation}
	{{\bf{E}}_s}{\bf{W}}\left( \nu \right) = {\bf{A}}\left( \nu \right){{\bf{B}}_s}\left( \nu \right) \;.
\label{eqn:WSF}
\end{equation}

In particular, $\bf{W} \left( {\nu } \right) $ must cancel the \emph{out of band} ${\bf{E}}_s$ components in the subspace ${{{\bf{A}}_ \bot }\left( \nu \right)}$
\begin{equation}
	{{\bf{B}}_{22}}\left( \nu \right){{\bf{C}}_{F2}}\left( \nu \right){\bf{W}}\left( \nu \right) = {\bf{0}}
	\label{eqn:outofbandcancellation}
\end{equation}
to cancel also the spectral leakage in the bin subspace ${{\bf{A}}\left( \nu \right)}$
\begin{equation}
	{{\bf{B}}_{12}}\left( \nu \right){{\bf{C}}_{F2}}\left( \nu \right){\bf{W}}\left( \nu \right) = {\bf{0}} \;.
	\label{eqn:inbandcancellation}
\end{equation}

All involved matrices can be rank deficient under certain conditions. In general, it is immediate to show that the signal subspace can be exactly recovered iff 
\begin{equation}
	\frac{{\left| {{{\bf{B}}_{12}}\left( \nu \right){{\bf{C}}_{F2}}\left( \nu \right){\bf{w}}} \right|_2^2}}{{\left| {{{\bf{B}}_{22}}\left( \nu \right){{\bf{C}}_{F2}}\left( \nu \right){\bf{w}}} \right|_2^2}} < \infty 
		\label{eqn:cancellationRayleighquotient}
\end{equation}
for any non-zero vector $\bf w$ which does not lie in the intersection of the null-spaces of ${\bf{B}}_{12}\left( {\nu } \right){{\bf{C}}_{F2}}\left( \nu \right)$ and ${\bf B}_{22}\left( {\nu} \right){\bf C}_{F2}\left( \nu \right)$\footnote{Intersection of these null spaces may not be empty for sources with spectral nulls within the array bandwidth, but in this case there is not any leakage to cancel. Infinite \eqref{eqn:cancellationRayleighquotient} indicates that a component does exist at frequency $\nu$ and not elsewhere, so it must be included in ${\bf{B}}_{11}\left( \nu \right)$.}.

Condition \eqref{eqn:cancellationRayleighquotient} just moves to the frequency domain the requirement of non-perfect coherency among narrow-band sources for the applicability of the spatial-only MUSIC \cite{SCHMIDT86} (i.e., spectral vs. spatial coherency), but cannot hamper the identifiability of spatially coherent sources from ${{\bf{B}}_s}\left( \nu \right)$. 

The structure of \eqref{eqn:columnspaceofEs} reveals that $\bf{W}$ can cancel out the spectral leakage from ${\bf{E}}_s $ separately for each uncorrelated signal, so ${{\bf{B}}_s}\left( \nu \right)$ contains linear combinations of all the steering vectors of active sources. Therefore, the \emph{limit} ST-MUSIC equation \cite{DICLAUDIO13,DICLAUDIO14},
\begin{equation}
	{\bf{E}}_v^H {\bf{A}}\left( \nu \right){{\bf{B}}_s}\left( \nu \right) = {\bf{0}} \;,
\label{eqn:ST-MUSIC}
\end{equation}
herein obtained by projecting both sides of \eqref{eqn:WSF} onto ${\bf{E}}_v$ is a \emph{sufficient} condition for the consistency of a signal subspace estimate.

DOA estimators starting from approximations of \eqref{eqn:ST-MUSIC} were considered in the past \cite{BUCKLEY88,YOON06,MOHAN08}. However, they employed non consistent, SCM based noise subspace formulations, basically unable to exploit the source power information and fruitfully deal with spatially coherent scenarios.

On the contrary, with consistent estimates of ${\bf{E}}_s$ and ${\bf{E}}_v$ for $N/MP \rightarrow\infty$, the sample ST-MUSIC subspace ${\hat{\bf{B}}_s}\left( \nu \right)$ asymptotically matches the \emph{ideal} narrow-band signal subspace model \cite{SCHMIDT86} and retains the full flexibility of the WSF approach in narrowband and broadband scenarios \cite{CADZOW90,VIBERG91,DICLAUDIO01} .

A different problem of ST-MUSIC is the unduly \emph{attenuation} of some sources of interest by \eqref{eqn:WSF}, evaluated as ${{\bf{B}}_{11}}\left( \nu \right){{\bf{C}}_{F1}}\left( \nu \right){\bf{W}}\left( \nu \right)$. In particular, harmonic sources made up by less than $L+P-1$ sinusoids with all frequencies different from $\nu$ or strongly cyclo-stationary sources with cycle frequency $\alpha  < {N^{ - 1}}$ \cite{GARDNER06} do not satisfy \eqref{eqn:ST-MUSIC} and might be utterly \emph{suppressed}. In fact, these sources have a rank deficient covariance block ${\bf S}_{dd}$ in \eqref{eqn:signalcovariancematrix}.

A basic ST-MUSIC subspace estimator was presented in \cite{DICLAUDIO13}, starting from a rank-reducing approach applied to ${\bf{E}}_s$, conceptually similar to TOPS \cite{YOON06}. An approximate ST-MUSIC for low SNR applications appeared in \cite{DICLAUDIO14}, based on the SVD of the weighted ${\bf{A}}{\left( \nu  \right)^H}{{\bf{E}}_s}$. These estimators used biased STCM estimates in finite samples and could not estimate $\kappa \left( \nu \right)$ in a statistically sound way. 

For these reasons, in the sequel we set up a \emph{ML inversion problem} of ${{\bf{B}}_s}\left( \nu \right)$, based on the first order (i.e., $O\left( {{N^{ - {1 \mathord{\left/
 {\vphantom {1 2}} \right.
 \kern-\nulldelimiterspace} 2}}}} \right)$) perturbation model \cite{GOLUB89,PILLAI89,VIBERG91,DICLAUDIO01} of the sample counterpart of \eqref{eqn:GEVD} 
\begin{equation}
	{\bf{R}}_v^{ - 1}{{\bf{\hat R}}_{ST}}{\bf{R}}_v^{ - H} = {\bf{\hat E\hat \Lambda }}{{\bf{\hat E}}^H}
	\label{eqn:sampleGEVD}
	\end{equation}
where
\begin{equation}
{{{\hat{\bf R}}}_{ST}} = \frac{1}{{N - P + 1}}\sum\limits_{n = P}^N {{{\bf{x}}_{ST}}\left( n \right){\bf{x}}_{ST}^H\left( n \right)} 
	\label{eqn:unbiasedSTCMestimate}
\end{equation}
is the \emph{unbiased} STCM estimate built from $N$ array output samples \eqref{eqn:wbsensorsignal}. 
 The sample eigenvalues of \eqref{eqn:sampleGEVD}, ${\hat \lambda _k} = {\bf{\hat \Lambda }}\left( {k,k} \right)$ are ordered for $k=1,2,\ldots,MP$ in a non-increasing manner. The corresponding orthonormal sample eigenvectors ${\hat{\bf E}}\left( {:,k} \right)$ are partitioned into the \emph{sample wide-band signal subspace} ${\hat{\bf E}}_s = {\hat{\bf E}}\left( {:,1:\eta} \right)$, of dimension $0 \le \eta < MP$, related to the dominant ${\hat \lambda _k}$, and the complementary \emph{sample wide-band noise subspace} ${\hat{\bf E}}_v = {\hat{\bf E}}\left( {:,\eta+1: MP} \right)$, related to the smallest $MP-\eta$ eigenvalues, clustered around $\lambda_v$ \cite{DICLAUDIO13,DICLAUDIO14}.
 
The EVD version of \eqref{eqn:sampleGEVD} is preferable since the statistical identification of a spherical ${\hat{\bf E}}_v$ ensures the feasibility of consistent DOA estimation and reduces the parametrization and the Mean Square Error (MSE) w.r.t. the true STCM \cite{LEDOIT04}. 

Under the hypotheses made in Sect. \ref{section:arraymodel}, two distinct STSs \eqref{eqn:spacetimesnapshot} ${{\bf{x}}_{ST}}\left( {{n_1}} \right)$ and ${{\bf{x}}_{ST}}\left( {{n_2}} \right)$ cannot be considered as statistically independent if $\left| {{n_1} - {n_2}} \right| \le P$. However, using independent STSs would require $N \gg MP^2$ to get a stable ${{\bf{\hat R}}_{ST}}$. 

In contrast, theoretical arguments \cite{PFAFFEL12} and past experience \cite{KROLIK90,DICLAUDIO13,DICLAUDIO14} show that \eqref{eqn:unbiasedSTCMestimate} converges only for $N \gg MP$ to its asymptotic performance, \emph{very similar} (except for leakage issues) to the one of a set of SCM estimates, using the same $N$ and a DFT length $P_w = P$, despite the much larger number of degrees of freedom of the STCM. The following analysis sheds light on this observation.

\section{ML ST-MUSIC Subspace Estimator}
\label{sec:STMUSICEstimator}
The present STCM analysis extends the classical one developed for the SCM \cite{GOLUB89,PILLAI89,VIBERG91,CHAMPAGNE94}, points out the differences and is validated by the simulations in Sect. \ref{sec:results}. In particular, it shows that the ML ST-MUSIC subspace departs from the sample version of \eqref{eqn:ST-MUSIC} \cite{DICLAUDIO13} and leads to a generalized eigen-problem.

\subsection{Sample STCM eigenvectors}
\label{sec:SampleSTCMEigenvectors}
The sample bases ${\hat{\bf E}}_s$ and ${\hat{\bf E}}_v$ are rotated versions of the true ones in \eqref{eqn:GEVD} \cite{GOLUB89} and converge to these for $N/MP\rightarrow\infty$ \cite{PILLAI89}. The (first-order) partitioned rotation matrix is defined by 
\begin{equation}
	\left[ {\begin{array}{*{20}{c}}
{{{{\hat{\bf E}}}_s}}&{{{{\hat{\bf E}}}_v}}
\end{array}} \right] \approx \left[ {\begin{array}{*{20}{c}}
{{{\bf{E}}_s}}&{{{\bf{E}}_v}}
\end{array}} \right]\left[ {\begin{array}{*{20}{c}}
{{\bf{I}} + {{\bf{G}}_{ss}}}&{{{\bf{G}}_{ns}}}\\
{{{\bf{G}}_{sn}}}&{{\bf{I}} + {{\bf{G}}_{nn}}}
\end{array}} \right]
\label{eqn:EVrotation}
\end{equation}
where ${\bf{G}}_{ns} = -{\bf{G}}_{sn}^H$, ${\bf{G}}_{ss} = -{\bf{G}}_{ss}^H$ and ${\bf{G}}_{nn} = -{\bf{G}}_{nn}^H$ are random \emph{perturbation derivatives} with $L_2$ norm of $O\left( {{\left| {{{\bf{R}}_{ST}}} \right|_2}{N^{ - {1 \mathord{\left/
 {\vphantom {1 2}} \right.
 \kern-\nulldelimiterspace} 2}}}} \right)$ \cite{PILLAI89}. 

The analysis is simplified by recognizing that blocks ${{\bf{I}} + {{\bf{G}}_{ss}}}$ and ${{\bf{I}} + {{\bf{G}}_{nn}}}$ are $O\left( {{N^{ - {1 \mathord{\left/
 {\vphantom {1 2}} \right.
 \kern-\nulldelimiterspace} 2}}}} \right)$ approximations to \emph{unitary} matrices \cite{MANTON02}. The invariant subspace ${\bf E}_v$ is intrinsically defined up to an arbitrary rotation. The rotation within ${{{{\hat{\bf E}}}_s}}$ was considered in random matrix theory \cite{MESTRE08} and covariance shrinkage \cite{LEDOIT04} and is strong for close signal eigenvalues. While the influence of this rotation on the accuracy bounds of DOA estimates is unknown, \eqref{eqn:ST-MUSIC} shows that \emph{full rank} linear mixing of ${\bf E}_s$ is unessential for estimator derivation, since a backward transformation is induced on ${\bf{W}}\left( \nu \right)$ by the numerical optimization on the particular realization of ${{{{\hat{\bf E}}}_s}}$. In addition, attempts to take into account ${{\bf{G}}_{ss}}$ may face undesired problems \cite{LEDOIT04}. 

After these observations, effects caused by ${\bf{G}}_{ss}$ and ${\bf{G}}_{nn}$ can be mostly neglected. Instead the random entries of ${{{\bf{G}}_{sn}}}$ are the most relevant for DOA estimation. They are the finite sample cosines between ${\hat{\bf E}}_s$ and the true ${\bf E}_v$, calculated as \cite{PILLAI89,GOLUB89,VIBERG91}
\begin{equation}
	{{\bf{G}}_{sn}}\left( {p,q} \right) = \frac{{{{\hat \rho }_{sn}\left( {p,q} \right)}}}{{{\lambda _q} - {\lambda _v}}}
	\label{eqn:Gsn}
\end{equation}
for $p=1,2,\ldots,MP-\eta$ and $q=1,2,\ldots,\eta$. In this equation, ${{{\hat \rho }_{sn}\left( {p,q} \right)}} = \frac{1}{{N - P + 1}}\sum\limits_{n = P}^{N} {{y_{\eta+p}}\left( n \right)y_q^*\left( n \right)}$ is the \emph{finite sample correlation} between the uncorrelated signals ${y_{\eta+p}}\left( n \right)$ and ${y_{q}}\left( n \right)$, where the generic ${y_k}\left( n \right) = {\bf{E}}{\left( {:,k} \right)^H}{\bf{R}}_v^{ - 1}{{\bf{x}}_{ST}}\left( n \right)$ is the signal \emph{extracted} by the transformed \emph{true} $k$-th eigenvector ${\bf{R}}_v^{ - H}{\bf{E}}\left( {:,k} \right)$, considered as a DS beamformer \cite{GINGRAS93} and referred to as \emph{eigenfilter}.

Therefore, $E\left\{ {{\hat \rho }_{sn}\left( {p,q} \right)} \right\} = 0+ O \left({N^{ - 1}}\right)$, while the covariances among ${\bf{G}}_{sn}$ entries are affected by the \emph{temporal correlation} between ${y_{\eta+p}}\left( n \right)$ and ${y_q}\left( n \right) $\cite{SANCETTA08}. For stationary signal and noise, the following covariances are computed after \cite{PILLAI89,MCCLOUD02}
\begin{equation}
	\begin{array}{c}
\begin{array}{*{20}{c}}
{{\rm{E}}\left\{ {{{\hat \rho }_{sn}}\left( {p,q} \right)\hat \rho _{sn}^*\left( {k,l} \right)} \right\} = }\\
{\sum\limits_{\tau = - \left( {N - P} \right)}^{N - P} {\frac{{N - P + 1 - \tau }}{{{{\left( {N - P + 1} \right)}^2}}}} {\rm{E}}\left\{ {{y_{\eta + p}}\left( n \right)y_q^*\left( n \right)} \right. \times }
\end{array}\\
\left. {y_{\eta + k}^*\left( {n + \tau } \right){y_l}\left( {n + \tau } \right)} \right\}
\end{array}
\label{eqn:gcovhermitian}
\end{equation}
\begin{equation}
	\begin{array}{c}
\begin{array}{*{20}{c}}
{{\rm{E}}\left\{ {{{\hat \rho }_{sn}}\left( {p,q} \right){{\hat \rho }_{sn}}\left( {k,l} \right)} \right\} = }\\
{\sum\limits_{\tau = - \left( {N - P} \right)}^{N - P} {\frac{{N - P + 1 - \left| \tau \right|}}{{{{\left( {N - P + 1} \right)}^2}}}} {\rm{E}}\left\{ {{y_{\eta + p}}\left( n \right)y_q^*\left( n \right) \times } \right.}
\end{array}\\
\left. {{y_{\eta + l}}\left( {n + \tau } \right)y_l^*\left( {n + \tau } \right)} \right\}\;.
\end{array}
\label{eqn:gcovstraight}
\end{equation}

With a large computational effort, some statistics of \eqref{eqn:gcovhermitian} and \eqref{eqn:gcovstraight} might be estimated from the sample ${y_q}\left( n \right) $. However, a great simplification can be obtained by examining the properties of STCM eigenfilters. 

In fact, the beampatterns of signal eigenfilters in the frequency/DOA space exhibit main-lobes pointing at the regions of activity of sources and extract \emph{convolutive mixtures} ${y_q}\left( n \right)$ ($q = 1,2, \ldots, \eta$) of source signals plus noise. 
\begin{figure}
	\centering
		\includegraphics[width=3.5in]{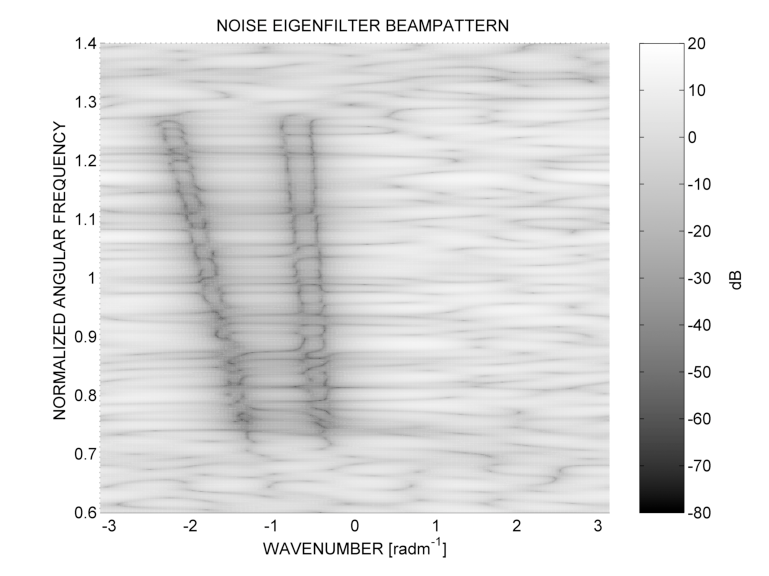}
		\includegraphics[width=3.5in]{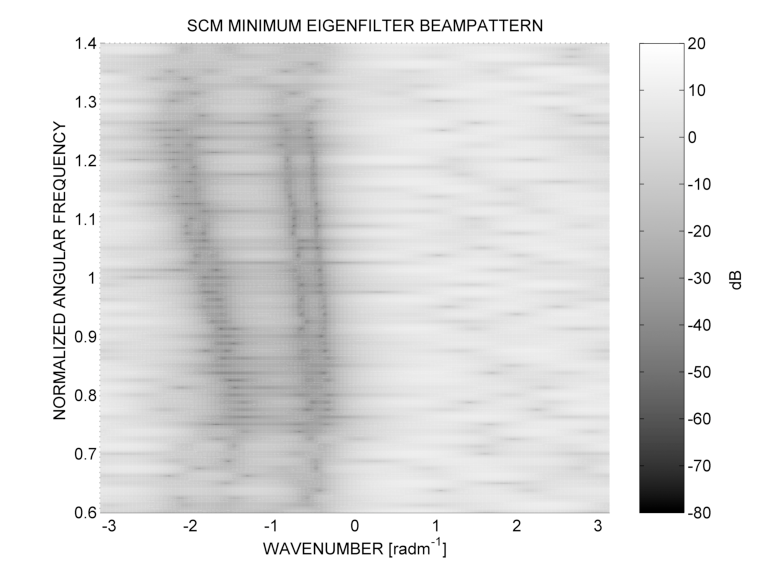}
	\caption{Top: Sample noise STCM eigenvector beampattern versus wavenumber and angular frequency, normalized to the array bandwidth center for four pass-band, uncorrelated sources, impinging on the ULA used for simulations in Sect. \ref{subsec:Widebandfocusing} for SNR$=20$ dB. Bottom: Average beampattern computed from the frequency-interpolated least dominant SCM eigenvectors of $64$ DFT bins from the same realization, showing null depth reduction and bias due to leakage artifacts. }
	\label{fig:Fig2}
\end{figure}

On the contrary, as shown in Fig. \ref{fig:Fig2}, which realizes Fig. \ref{fig:Fig1} in a computer simulation, noise eigenfilters generally have beampatterns with very deep broad-band nulls steered toward the source DOAs \cite{BUCKLEY88}. Therefore, they extract a set of noise dominated signals ${y_{\eta+p}}\left( n \right)$ ($p = 1,2, \ldots, MP - \eta$) that are \emph{almost white and mutually uncorrelated} and additionally uncorrelated with any source dominated signal ${y_{q}}\left( n \right)$ ($q = 1,2, \ldots, \eta$), because \eqref{eqn:sampleGEVD} removes in a Least Squares (LS) sense \cite{KAYMARPLE81} the temporal correlation from ${y_{\eta+p}}\left( n \right)$ for any $\tau < P$.
 
Finally, if signals and noise are realizations of zero mean, circularly complex processes for every $\tau$ with zero third-order moments, it is possible to approximate \eqref{eqn:gcovhermitian} and \eqref{eqn:gcovstraight} as 
\begin{equation}
	{\rm{E}}\left\{ {{{\hat \rho }_{sn}}\left( {p,q} \right)\hat \rho _{sn}^*\left( {k,l} \right)} \right\} \approx {\left( {N - P + 1} \right)^{ - 1}}{\delta _{pk}}{\delta _{ql}}{\lambda _v}{\lambda _q}
\label{eqn:gcovhermitianapprox}
\end{equation}
\begin{equation}
	{\rm{E}}\left\{ {{{\hat \rho }_{sn}}\left( {p,q} \right){{\hat \rho }_{sn}}\left( {k,l} \right)} \right\} = 0
\label{eqn:gcovapprox}
\end{equation}
independently of the exact distributions of signals and noise \cite{PILLAI89}. For $N \gg MP$, the entries of ${\bf{G}}_{sn}$ approach a Gaussian distribution by a Central Limit Theorem argument \cite{VIBERG91}. Then asymptotically \eqref{eqn:Gsn} can be approximated as 
\begin{equation}
{{\bf{G}}_{sn}} \approx {{\bf{G}}_0}{\bf{\Gamma }}_0^{{1 \mathord{\left/
 {\vphantom {1 2}} \right.
 \kern-\nulldelimiterspace} 2}}
\label{eqn:Gsnapprox}
\end{equation}
where ${\bf{G}}_0$ is a $\left( {MP-\eta}\right) \times \eta$ random matrix with i.i.d., zero mean circular Gaussian entries of variance ${\left( {N - P + 1} \right)^{ - 1}}$ and
\begin{equation}
	{\bf{\Gamma }}_0^{1/2} = {\rm{diag}}\left\{ {\left[ {\begin{array}{*{20}{c}}
{\gamma _0^{1/2}\left( 1 \right)}& \cdots &{\gamma _0^{1/2}\left( \eta \right)}
\end{array}} \right]} \right\}
	\label{eqn:Gamma0}
\end{equation}
where ${\gamma _0}\left( k \right) = {\lambda _k}{\lambda _v}{\left( {{\lambda _k} - {\lambda _v}} \right)^{ - 2}}$, exactly like in the narrow-band case using $N-P+1$ independent snapshots \cite{VIBERG91}. 

Because of the random rotations ${{\bf{G}}_{ss}}$ and ${{\bf{G}}_{nn}}$, ${\bf{G}}_0$ is not observable, but filtered versions of it enter the analysis as large random Wigner covariance matrices (e.g., ${\bf{G}}_0^H{{\bf{G}}_0}$). Wigner matrices converge to a common limit distribution with Gaussian off-diagonal entries and a \emph{semicircle} eigenvalue distribution, regardless reasonable violations of \eqref{eqn:gcovhermitianapprox} and \eqref{eqn:gcovapprox} \cite{HOFMANN08}. This observation well explains the empirical results of \cite{KROLIK90,DICLAUDIO13,DICLAUDIO14}.

Curiously, the largest deviations are expected for very small $ \gamma_0 \left( k \right) \approx {\lambda _v}/{\lambda _k}$, i.e., at \emph{high SNR}, and in the presence of strong temporal correlation of source signals.

\subsection{Normalized signal eigenvectors}
\label{sec:NormalizedSignaleigenvectors}
Perturbed signal eigenvectors in \eqref{eqn:EVrotation} are orthonormal only up to $O\left( {{N^{ - {1 \mathord{\left/
 {\vphantom {1 2}} \right.
 \kern-\nulldelimiterspace} 2}}}} \right)$ terms \cite{PILLAI89}. From \eqref{eqn:Gamma0}, these eigenvectors exhibit a non-physical singularity for ${\lambda _k} \to {\lambda _v}$, when sample unit norm eigenvectors only slip in the estimated $\hat{\bf E}_v$. To circumvent this problem, in the sequel we will normalize the $k$-th perturbed signal eigenvector \eqref{eqn:EVrotation} with respect to its expected $L_2 $ norm $\sqrt {1 + c{\gamma _0}\left( k \right)} $, where $c = {{\left( {MP - \eta } \right)} \mathord{\left/
 {\vphantom {{\left( {MP - \eta } \right)} {\left( {N - P + 1} \right)}}} \right.
 \kern-\nulldelimiterspace} {\left( {N - P + 1} \right)}}$ is the ratio between the dimension of ${\bf E}_v$ and the number of STSs used in \eqref{eqn:unbiasedSTCMestimate} and resembles a parameter defined in random matrix theory \cite{MESTRE08,NADAKUDITI08}. 

This \emph{fixed} $O\left( {{N^{ - 1}}} \right)$ scaling cannot modify the asymptotic subspace performance for $N\rightarrow\infty$ \cite{PILLAI89}. Even if perturbations do not likely follow the first order modeling for ${\lambda _k} \to {\lambda _v}$, this scaling brings out correction terms that taper off the influence of signal eigenvalues down to zero in the same limit and partially \emph{move} marginal eigenvectors into $\hat{\bf E}_v$. However, the corrections are not negligible for typical values of $N$ and ${\bf{\Gamma }}$, therefore increasing the robustness to estimation errors of $\eta$. 

The scaled perturbation model \eqref{eqn:EVrotation} of $\hat{\bf E}_s$, neglecting the inessential rotation induced by ${{\bf{G}}_{ss}}$, becomes
\begin{equation}
	{{{\hat{\bf E}}}_s} \approx {{\bf{E}}_s}{\bm{\Gamma }}_s^{{1 \mathord{\left/
 {\vphantom {1 2}} \right.
 \kern-\nulldelimiterspace} 2}} + {{\bf{E}}_v}{{\bf{G}}_0}{{\bm{\Gamma }}^{{1 \mathord{\left/
 {\vphantom {1 2}} \right.
 \kern-\nulldelimiterspace} 2}}}
\label{eqn:perturbedWSS}
\end{equation}
where ${\bf{\Gamma }}_s^{{1 \mathord{\left/
 {\vphantom {1 2}} \right.
 \kern-\nulldelimiterspace} 2}}$ and ${{\bf{\Gamma }}^{1/2}}$ are the diagonal square roots \cite{GOLUB89} of ${{\bf{\Gamma }}_s} = {\left[ {{{\bf{I}}_\eta } + c{{\bf{\Gamma }}_0}} \right]^{ - 1}}$ and ${\bf{\Gamma }} = {{\bf{\Gamma }}_0}{\left[ {{{\bf{I}}_\eta } + c{{\bf{\Gamma }}_0}} \right]^{ - 1}}$. In addition, $0 < c{\bf{\Gamma }}\left({ k,k }\right) < 1$ and \eqref{eqn:Gsnapprox} can be recovered for $c\rightarrow 0$. 

\subsection{Signal subspace rank estimation for the STCM}
\label{subsec:SignalSubspaceRankEstimationForTheSTCM}

From the previous arguments, it is clear that the rank $\eta$ of ${\bf E}_s$ is not directly related to the number of uncorrelated impinging signals, as in the narrow-band case \cite{WAX85}, but it represents the number of STCM signal components that exceed the noise level $\lambda_v$ and carry significant information about source parameters. In essence, the problem is to find a spherical noise subspace centered on a consistent $O\left( {{N^{ - {1 \mathord{\left/
 {\vphantom {1 2}} \right.
 \kern-\nulldelimiterspace} 2}}}} \right)$ estimate of $\lambda_v$. This goal involves the STCM eigenvalue statistics, still affected by the snapshot dependency issue \cite{SANCETTA08}. 

Under the same assumptions made in Sect. \ref{sec:SampleSTCMEigenvectors}, the perturbative model for the sample STCM eigenvalues $\hat {\lambda} _k$ starts from the approximation ${\hat \lambda _k} \approx {\left( {N - P + 1} \right)^{ - 1}}\sum\limits_{n = P}^N {{{\left| {{y_k}\left( n \right)} \right|}^2}} $\cite{GOLUB89,PILLAI89}. It is deduced that ${\mathop{\rm E}\nolimits} \left\{ {{{\hat \lambda }_k}} \right\} = {\lambda _k} + O \left( {N^{ - 1}} \right)$ for $k=1,2,\ldots,\eta$ and ${\mathop{\rm E}\nolimits} \left\{ {{\hat {\lambda }_k}} \right\} = {\lambda _v} + O \left( {N^{ - 1}} \right)$ for $k=\eta+1,\eta+2,\ldots,MP$. 

For large $N$, the covariance between ${\hat \lambda }_k$ and ${\hat \lambda }_l$, corresponding to \emph{distinct} $\lambda_k \neq \lambda_l$, as well as the variance of \emph{isolated} signal $\hat {\lambda} _k$, approach \cite{PILLAI89,SANCETTA08}
\begin{equation}
	\begin{array}{c}
{\mathop{\rm cov}} \left( {{{\hat \lambda }_k},{{\hat \lambda }_l}} \right) \simeq \sum\limits_{\tau = - \left( {N - P} \right)}^{N - P} {\frac{{N - P + 1 - \left| \tau \right|}}{{{{\left( {N - P + 1} \right)}^2}}}} \times \\
{\mathop{\rm E}\nolimits} \left\{ {{{\left| {{y_k}\left( n \right)} \right|}^2}{{\left| {{y_l}\left( {n + \tau } \right)} \right|}^2}} \right\} - {\lambda _k}{\lambda _l} \;.
\end{array}
\label{eqn:eigenvaluecov}
\end{equation}

In particular, pairs of signal and noise sample eigenvalues are almost uncorrelated. However, \eqref{eqn:eigenvaluecov} indicates that the variance of signal STCM eigenvalues is much higher in the dependent snapshot case, but still generally lower than in the SCM case. 

As regards the sample noise eigenvalue statistics, the consistency of the classical noise variance estimate \cite{WAX85}
\begin{equation}
	{\hat \lambda _v} = \frac{1}{{MP - \eta }}\sum\limits_{k = \eta + 1}^{MP - \eta } {{{\hat \lambda }_k}} 
	\label{eqn:samplenoisevariance}
\end{equation}
follows from the convergence of \eqref{eqn:unbiasedSTCMestimate} to the true STCM under the given assumptions on the choice of $P$ \cite{PFAFFEL12}. 

In addition, the robust censoring estimator of $\eta$ proposed in \cite{DICLAUDIO01} showed that the sample \emph{ensemble variance} ${\mathop{\rm var}} \left( {\left\{ {{{\hat \lambda }_{\eta + p};\,p=1,2,\ldots,MP-\eta}} \right\}} \right)$ of \eqref{eqn:sampleGEVD} is always close to $ c\lambda _v^2$ for Gaussian noise, as in the independent snapshot case \cite{NADAKUDITI08}. This result confirms that the extracted noise signals ${y_{\eta + p}}\left( n \right)$ behave as white and mutually uncorrelated for any $\tau$ of interest, in agreement with the analysis of \cite{PFAFFEL12} for more general noise distributions. However, a small increase of this ensemble variance was observed at low SNR and was interpreted as a symptom of increasing leakage of temporally correlated signals into $\hat{\bf E}_v$ .

Therefore, by Chebyschev's inequality, the lower threshold for signal eigenvalues should be a small multiple of $\sqrt c {\hat \lambda _v}$ above ${\hat \lambda _v}$ to avoid missing the small components originated from the fading tails of the impulse responses ${{\bf{h}}_m}\left( {{{\bm{\theta }}_d}} \right)$\cite{DICLAUDIO13}.  

It follows that any optimal value of $\eta$ shrinks with the SNR and this phenomenon may impact on the DOA identifiability by \eqref{eqn:ST-MUSIC}, since a reduced dimension ${\bf{W}}\left( \nu \right)$ may not consistently recover the full span of ${\bf B}_s \left( \nu \right)$ anymore. 

Information Theoretic criteria \cite{WAX85,NADAKUDITI08} developed for the SCM are questionable for the STCM, because signal and noise processes exhibit very different and a priori unknown numbers of effective observations by \eqref{eqn:eigenvaluecov} and the likelihood form is different for each signal distribution. 

However, the Gaussian log likelihood ratio \cite{WAX85} for independent observations essentially relies on the noise eigenvalue spread, that changes little under the assumption of Sect. \ref{sec:SampleSTCMEigenvectors}, even for many non-Gaussian noise distributions \cite{PFAFFEL12}. These arguments support the provisional applicability of unmodified existing SCM rank estimation criteria to the STCM. In a refined model, the dependent observation issues might be handled by assuming a smaller $\bar{N} < N-P+1$ number of observations \cite{SANCETTA08} in the relevant noise subspace, but a quantitative study is beyond the scope of this work. 

In particular, the AIC \cite{WAX85}, which takes into account only the validation risk \cite{BURNHAM04} of adding a new signal eigenvector, performed better, especially at low SNR, in comparison with the non parametric approach of \cite{DICLAUDIO01}, the BIC \cite{WAX85} and the modified AIC of \cite{NADAKUDITI08}, that evidenced clear mismatches of their additional parameters. In particular, despite of consistency claims and stable $\eta$ estimation at high SNR, the BIC exhibited a disappointing detection threshold at low SNR, about $10$ dB higher than the AIC one. Therefore, only the AIC will be included in the simulations of Sect. \ref{sec:results}.

\subsection{ML ST-MUSIC inversion}
\label{sec:MLSTMUSICinversion}
The main issue with earlier versions of ST-MUSIC \cite{DICLAUDIO13,DICLAUDIO14}, based on the SVD of ${\bf{\hat E}}_s^H{\bf{A}}\left( \nu \right)$, was the statistical instability of the signal subspace weighting, which generalized existing weighted MUSIC schemes \cite{STOICA90a,MCCLOUD02}. In particular, the empirical estimate ${\bf{\hat \Gamma }}$ of the large $\bf \Gamma$ led to signal cancellation at high SNR, as in adaptive beamforming \cite{LI05}. Heuristic regularization of ${\bf{\hat \Gamma }}$, based on \cite{LEDOIT04}, stabilized the subspace estimate at the expense of the sample DOA bias, partially spoiling the goal of consistency. 

A true ML formulation circumvented these issues by rewriting the sample version of \eqref{eqn:WSF} at frequency $\nu$, after inserting \eqref{eqn:perturbedWSS}, as 
\begin{equation}
	{{\bf{\hat E}}_s}{\bf{W}}\left( \nu  \right) = {\bf{A}}\left( \nu  \right){\bf{B}}\left( \nu  \right) + {{\bf{E}}_v}{{\bf{G}}_0}{{\bf{\Gamma }}^{{1 \mathord{\left/
 {\vphantom {1 2}} \right.
 \kern-\nulldelimiterspace} 2}}}{\bf{W}}\left( \nu  \right) + o\left( {{N^{ - {1 \mathord{\left/
 {\vphantom {1 2}} \right.
 \kern-\nulldelimiterspace} 2}}}} \right)
\label{eqn:empiricalSTMUSIC}
\end{equation}
where ${{\bf{B}}}\left( \nu \right)$ is the unknown \emph{candidate signal subspace basis} at $\left( \nu \right)$ for ${{\bf{B}}_s}\left( \nu \right)$ in \eqref{eqn:narrowbandWSF}. 

It is assumed that $\eta$ has been estimated as described in Sect.\ref{subsec:SignalSubspaceRankEstimationForTheSTCM}. The true $ {{\bf{\Gamma}}}$ and ${\bf{W}}\left( \nu \right) $ are assumed initially known. In the sequel, for conciseness we will set $ N_P = N-P+1 $ and drop the reference to $\nu$, e.g., $\bf A = {\bf{A}}\left( \nu \right)$, $\bf B = {\bf{B}}\left( \nu \right)$ and $\bf W = {\bf{W}}\left( \nu \right)$. 

The sample \emph{equation error} ${\bf E}_v{{\bf{G}}_0}{{\bf{\Gamma }}^{{1 \mathord{\left/
 {\vphantom {1 2}} \right.
 \kern-\nulldelimiterspace} 2}}}{\bf{W}}$ of \eqref{eqn:empiricalSTMUSIC} lies in the \emph{true} ${{{{\bf E}}}_v}$. One of the instability sources was found in the classical replacement of ${{{{\bf E}}}_v}$ by the sample ${{{\hat{\bf E}}}_v}$, that called for a more sophisticated route. Projecting \eqref{eqn:empiricalSTMUSIC} onto ${{{\hat{\bf E}}}_v}$ and ${{{\hat{\bf E}}}_s}$ leads to
\begin{equation}
{\hat{\bf E}}_v^H{{\bf{E}}_v}{{\bf{G}}_0}{{\bf{\Gamma }}^{{1 \mathord{\left/
 {\vphantom {1 2}} \right.
 \kern-\nulldelimiterspace} 2}}}{\bf{W}} = {\hat{\bf E}}_v^H{\bf{A{{\bf{B}}}}} + o\left( {{N^{ - {1 \mathord{\left/
 {\vphantom {1 2}} \right.
 \kern-\nulldelimiterspace} 2}}}} \right) 
\label{eqn:GLSequationError}
\end{equation}
\begin{equation}
\left( {{{\bf{I}}_\eta } - {{\bm{\Gamma }}^{{1 \mathord{\left/
 {\vphantom {1 2}} \right.
 \kern-\nulldelimiterspace} 2}}}{\bf{G}}_0^H{{\bf{G}}_0}{{\bm{\Gamma }}^{{1 \mathord{\left/
 {\vphantom {1 2}} \right.
 \kern-\nulldelimiterspace} 2}}}} \right){\bf{W}} = {\hat{\bf E}}_s^H{\bf{A{{\bf{B}}}}} + o\left( {{N^{ - {1 \mathord{\left/
 {\vphantom {1 2}} \right.
 \kern-\nulldelimiterspace} 2}}}} \right) 
\label{eqn:GLSequationWeight}
\end{equation}
and
\begin{equation}
	{\hat{\bf E}}_s^H{{\bf{E}}_v}{{\bf{G}}_0}{{\bm{\Gamma }}^{{1 \mathord{\left/
 {\vphantom {1 2}} \right.
 \kern-\nulldelimiterspace} 2}}}{\bf{W}} = {{\bm{\Gamma }}^{{1 \mathord{\left/
 {\vphantom {1 2}} \right.
 \kern-\nulldelimiterspace} 2}}}{\bf{G}}_0^H{{\bf{G}}_0}{{\bm{\Gamma }}^{{1 \mathord{\left/
 {\vphantom {1 2}} \right.
 \kern-\nulldelimiterspace} 2}}}{\bf{W}} + o\left( {{N^{ - {1 \mathord{\left/
 {\vphantom {1 2}} \right.
 \kern-\nulldelimiterspace} 2}}}} \right) \;.
\label{eqn:ErrorInSampleWSS}
\end{equation}

Equations \eqref{eqn:GLSequationError} and \eqref{eqn:ErrorInSampleWSS} together characterize a \emph{row rotation} of ${\bf E}_v{{\bf{G}}_0}{{\bf{\Gamma }}^{{1 \mathord{\left/
 {\vphantom {1 2}} \right.
 \kern-\nulldelimiterspace} 2}}}{\bf{W}}$, while \eqref{eqn:GLSequationWeight} can be used to estimate ${\bf{W}}$. After \eqref{eqn:Gsnapprox}, ${{\bf{G}}_0}{{\bm{\Gamma }}^{{1 \mathord{\left/
 {\vphantom {1 2}} \right.
 \kern-\nulldelimiterspace} 2}}}{\bf{W}}$ can be modeled without loss of generality as a stack of $MP -\eta$ independent, zero mean, Gaussian circular observations of dimension $M$ with covariance $N_P^{ - 1}{{\bf{W}}^H}{\bm{\Gamma W}}$, characterized by the following negative log likelihood for given ${\bf{W}}$ and ${\bm{\Gamma }}$
\begin{equation}
	\begin{array}{c}
\varphi\left( M \right) = M\left( {MP - \eta } \right)\ln \left( \pi \right) + \\
\left( {MP - \eta } \right)\ln \det \left( {N_P^{ - 1}{{\bf{W}}^H}{\bf{\Gamma W}}} \right) + \\
{N_P}\,{\mathop{\rm trace}\nolimits} \left[ {{{\bf{\Pi}}_\varepsilon^{(0)} }{{\left( {{{\bf{W}}^H}{\bf{\Gamma W}}} \right)}^{ - 1}}} \right]
\end{array}
\label{eqn:loglikelihood0}
\end{equation}
where \begin{displaymath}
\begin{array}{c}
{{\bf{\Pi}}_\varepsilon^{(0)} } = {\bf{B}}^H{{\bf{A}}^H}{{{\bf{\hat E}}}_v}{\bf{\hat E}}_v^H{\bf{A}}{{\bf{B}}} + \\
{{\bf{W}}^H}{{\bf{\Gamma }}^{{1 \mathord{\left/
 {\vphantom {1 2}} \right.
 \kern-\nulldelimiterspace} 2}}}{\bf{G}}_0^H{{\bf{G}}_0}{\bf{\Gamma G}}_0^H{{\bf{G}}_0}{{\bf{\Gamma }}^{{1 \mathord{\left/
 {\vphantom {1 2}} \right.
 \kern-\nulldelimiterspace} 2}}}{\bf{W}}\;{\kern 1pt} 
\end{array}\;.
\label{eqn:equationerror0}
\end{displaymath}

The last term of ${{\bf{\Pi}}_\varepsilon^{(0)} }$ can be replaced by its expected value under the Gaussian i.i.d. approximation on ${\bf G}_0$, leading to
\begin{equation}
	 \begin{array}{c}
{{\bf{\Pi}}_\varepsilon^{(0)} } = {{{\bf{B}}}^H}{{\bf{A}}^H}{{{\hat{\bf E}}}_v}{\hat{\bf E}}_v^H{\bf{AB}} + \\
{{\bf{W}}^H}\left[ {{c^2}{{\bf{\Gamma }}^2} + c{\bm{\Gamma }} \cdot N_P^{ - 1}{\mathop{\rm trace}\nolimits} \left( {\bm{\Gamma }} \right)} \right]{\bf{W}} \;.
\end{array}
\label{eqn:equationerror1}
\end{equation}

The term ${\left( {{{\bf{W}}^H}{\bf{\Gamma W}}} \right)^{ - 1}}$ in \eqref{eqn:loglikelihood0} can be interpreted as the high resolution \emph{Capon estimate} \cite{CAPON69} of the power spectral density (PSD) at frequency $\nu$ of a set of signals characterized by the covariance ${{\bm{\Gamma }}^{ - 1}}$ and the steering vector matrix ${\bf{W}}$. As a consequence, signals components at frequencies different from $\nu$, such as spectral leakage products, are suppressed as in MV beamforming \cite{LI05}. The same term in \eqref{eqn:equationerror1} compensates for the inclusion of marginal signal eigenvalues close to ${\hat \lambda _v}$.

The $O\left( {{N^{ - 2}}} \right)$ term ${c^2}{{\bf{W}}^H}{{\bf{\Gamma }}^2}{\bf{W}}$ in \eqref{eqn:equationerror1} cannot influence the asymptotic performance, but it is the key for regularizing \eqref{eqn:loglikelihood0} at any SNR by filling the numerical rank of ${{\bf{\Pi}}_\varepsilon^{(0)} }$and compensating for any rotation of the sample $\bf W$. Its exact scaling mildly depends on the assumed distribution of ${\bf{G}}_0^H{{\bf{G}}_0}$, but the given calculus for the Gaussian case is sufficient for regularization purposes.

Plugging \eqref{eqn:equationerror1} into \eqref{eqn:loglikelihood0} and simplifying leads to 
\begin{equation}
	\begin{array}{c}
\varphi \left( M \right) = M\left( {MP - \eta } \right)\ln \left( {\pi N_P^{ - 1}} \right) + \\
\left( {MP - \eta } \right)\ln \det \left( {{{\bf{W}}^H}{\bm{\Gamma W}}} \right) + cM{\mathop{\rm trace}\nolimits} \left( {\bf{\Gamma }} \right) + {N_P} \times \\
{\mathop{\rm trace}\nolimits} \left[ {\left( {{{{\bf{B}}}^H}{{\bf{A}}^H}{{{\hat{\bf E}}}_v}{\hat{\bf E}}_v^H{\bf{AB}} + {c^2}{{\bf{W}}^H}{{\bm{\Gamma }}^2}{\bf{W}}} \right){{\left( {{{\bf{W}}^H}{\bm{\Gamma W}}} \right)}^{ - 1}}} \right] \;.
\end{array}
\label{eqn:loglikelihood1}
\end{equation}

There is some freedom in choosing an estimate ${\hat {\bf{W}}}$ of ${\bf{W}}$ from \eqref{eqn:GLSequationWeight}. Early weighted forms of MUSIC DOA estimators \cite{STOICA90a,MCCLOUD02} obtained null spectra resembling the last term of \eqref{eqn:loglikelihood0} for the implicit choice ${\hat {\bf{W}}} = {\hat{\bf E}}_s^H{\bf{AB}}$. However, errors in ${\hat {\bf{W}}}$ may destroy the Capon spectral estimate of \eqref{eqn:loglikelihood1} \cite{LI05}. Therefore, we sought for linear estimates of ${ {\bf{W}}}$, independent of ${\bf G}_0$, that minimize both the residual error of \eqref{eqn:GLSequationWeight} and the variance of ${\hat {\bf{W}}}$, according to \eqref{eqn:perturbedWSS}. In particular, the form of the random term of \eqref{eqn:perturbedWSS} suggested to minimize the expected value over ${\bf G}_0$ of the weighted LS functional
\begin{displaymath}	
 {\left| {{{\bm{\Gamma }}^{{{ - 1} \mathord{\left/
 {\vphantom {{ - 1} 2}} \right.
 \kern-\nulldelimiterspace} 2}}}\left[ {{\hat{\bf E}}_s^H{\bf{AB}} - \left( {{{\bf{I}}_\eta } - {{\bm{\Gamma }}^{{1 \mathord{\left/
 {\vphantom {1 2}} \right.
 \kern-\nulldelimiterspace} 2}}}{\bf{G}}_0^H{{\bf{G}}_0}{{\bm{\Gamma }}^{{1 \mathord{\left/
 {\vphantom {1 2}} \right.
 \kern-\nulldelimiterspace} 2}}}} \right){\bf{W}}} \right]} \right|_F^2} 
\end{displaymath}	
w.r.t. $\bf W$, leading to the asymptotically unbiased estimate
\begin{equation}
	\begin{array}{c}
{\hat {\bf{W}}} = {\bf{\Phi \hat E}}_s^H{\bf{A{{\bf{B}}}}} 
\end{array}
\label{eqn:WLSweight}
\end{equation}
where ${\bf{\Phi }} = {\left[ {{{\bf{I}}_\eta } - 2c{\bf{\Gamma }} + \left( {{c^2} + c\eta N_P^{ - 1}} \right){{\bf{\Gamma }}^2}} \right]^{ - 1}}\left( {{{\bf{I}}_\eta } - c{\bf{\Gamma }}} \right)$ is a real-valued, diagonal matrix of size $\eta$.

The candidate solution for ${\bf{B}}$ can be found from \eqref{eqn:loglikelihood1} through the \emph{Generalized SVD} (GSVD) \cite{GOLUB89}, which leads in a square root fashion to the decompositions
\begin{equation}
{\bm \Pi}_\varepsilon = {{\bf{A}}^H}{{{\hat{\bf E}}}_v}{\hat{\bf E}}_v^H{\bf{A}} + {c^2}{{\bf{A}}^H}{{{\hat{\bf E}}}_s}{\bf{\Phi }}{{\bf{\Gamma }}^2}{\bf{\Phi \hat E}}_s^H{\bf{A}} = {\bf{F\Sigma }}_\varepsilon^2{{\bf{F}}^H}
\label{eqn:GSVD1}
\end{equation}
\begin{equation}
{\bm \Pi}_s = {{\bf{A}}^H}{{{\hat{\bf E}}}_s}{\bf{\Phi \Gamma \Phi \hat E}}_s^H{\bf{A}} = {\bf{F\Sigma }}_s^2{{\bf{F}}^H}
\label{eqn:GSVD2}
\end{equation}
where ${\bm{\Sigma }}_s^2$ and ${\bm{\Sigma }}_\varepsilon^2$ are diagonal square matrices of size $M$ with non-negative diagonal entries, linked by ${\bm{\Sigma }}_\varepsilon^2 + {\bm{\Sigma }}_s^2 = {{\bf{I}}_M}$, and $\bf F$ is a complex valued square matrix of size $M$. By posing ${\bf{B}} = {{\bf{F}}^{ - H}}$ \cite{GOLUB89}, \eqref{eqn:loglikelihood1} can be simplified as
\begin{equation}
	\begin{array}{c}
{\varphi \left(M \right)} = M\left( {MP - \eta } \right)\ln \left( {\pi N_P^{ - 1}} \right) + \\
\left( {MP - \eta } \right)\ln \det \left( {{\bm{\Sigma }}_s^2} \right) + cM{\mathop{\rm trace}\nolimits} \left( {\bf{\Gamma }} \right) + \\
{N_P}{\mathop{\rm trace}\nolimits} \left( {{\bm{\Sigma }}_\varepsilon^2{\bm{\Sigma }}_s^{ - 2}} \right) \;.
\end{array} 
\label{eqn:likelihood2}
\end{equation}

Not every column of ${\bf{B}}$ is acceptable as a basis vector for the signal subspace, since $\varphi \left( M \right)$ has small contributes only along the $\kappa$ columns of ${\bf{B}}$ characterized by very small generalized eigenvalues ${\mu _k} = {{{\bm{\Sigma }}_\varepsilon^2\left( {k,k} \right)} \mathord{\left/
 {\vphantom {{{\bm{\Sigma }}_\varepsilon^2\left( {k,k} \right)} {{\bm{\Sigma }}_s^2\left( {k,k} \right)}}} \right.
 \kern-\nulldelimiterspace} {{\bm{\Sigma }}_s^2\left( {k,k} \right)}} \simeq c$. 

In fact, plugging \eqref{eqn:perturbedWSS} into \eqref{eqn:GSVD1} and \eqref{eqn:GSVD2} and taking the expected value over ${\bf G}_0$, leads to the following $O\left( {{N^{ - 1}}} \right)$ approximations
\begin{equation}
\begin{array}{c}
{\mathop{\rm E}\nolimits} \left\{ {{{\bf{\Pi }}_\varepsilon }} \right\} \simeq {{\bf{I}}_M} - {\mathop{\rm E}\nolimits} \left\{ {{{\bf{A}}^H}{{{\bf{\hat E}}}_s}\left[ {{{\bf{I}}_\eta } - {c^2}{\bf{\Phi }}{{\bf{\Gamma }}^2}{\bf{\Phi }}} \right]{\bf{\hat E}}_s^H{\bf{A}}} \right\}\\
 \simeq {{\bf{A}}^H}{{\bf{E}}_s}c{\bf{\Gamma }}\left( {{{\bf{I}}_\eta } + c{\bf{\Gamma }}} \right){\bf{E}}_s^H{\bf{A}} + \\
\left[ {1 - N_P^{ - 1}{\mathop{\rm trace}\nolimits} \left( {{\bf{\Gamma }} - {c^2}{{\bf{\Gamma }}^3}{{\bf{\Phi }}^2}} \right)} \right]{{\bf{A}}^H}{{\bf{E}}_v}{\bf{E}}_v^H{\bf{A}} 
\end{array}
\label{eqn:GSVDas1}
\end{equation}
\begin{equation}
\begin{array}{c}
{\mathop{\rm E}\nolimits} \left\{ {{{\bf{\Pi }}_s}} \right\} \simeq {{\bf{A}}^H}{{\bf{E}}_s}{\bf{\Gamma }}\left( {{{\bf{I}}_\eta } + c{\bf{\Gamma }}} \right){\bf{E}}_s^H{\bf{A}} + \\
N_P^{ - 1}{\mathop{\rm trace}\nolimits} \left( {{{\bf{\Gamma }}^2}{{\bf{\Phi }}^2}} \right){{\bf{A}}^H}{{\bf{E}}_v}{\bf{E}}_v^H{\bf{A}}
\end{array}
\label{eqn:GSVDas2}
\end{equation}
where some non-negligible terms from the Taylor series expansions of \eqref{eqn:WLSweight}, \eqref{eqn:GSVD1} and \eqref{eqn:GSVD2} w.r.t. $c$ are retained for clarity and future use. 

Thus, on the average, for $N\rightarrow\infty$ the generalized eigen-equation corresponding to \eqref{eqn:GSVD1} and \eqref{eqn:GSVD2} \cite{GOLUB89}
\begin{equation}
	{\rm{E}}\left\{ {{{\bf{\Pi }}_\varepsilon }} \right\}{{\bf{B}}\left( {:,k} \right)} \simeq {\mu _k}{\mkern 1mu} {\rm{E}}\left\{ {{{\bf{\Pi }}_s}} \right\}{{\bf{B}}\left( {:,k} \right)}
	\label{eqn:signalsubspaceeigenvalue}
\end{equation}
is asymptotically satisfied by an eigenvalue $\mu_k = c$ of multiplicity $\kappa$. The corresponding $\kappa$ eigenvectors, collected in the $M \times\ \kappa$ matrix 
${{\bf{B}}_s}$, asymptotically satisfy ${{\bf{A}}^H}{{\bf{E}}_v}{\bf{E}}_v^H{\bf{A}} {{{\bf{B}}_s}} \simeq {\bf{0}}$ and therefore \eqref{eqn:ST-MUSIC}. 

In addition, the null-space of ${\rm{E}}\left\{ {{{\bf{\Pi }}_\varepsilon } - c{{\bf{\Pi }}_s}} \right\}$ is independent of $N$ for sufficiently large samples, demonstrating that the \emph{span} of the sample ${{\bf{B}}_s}$ (i.e., $\hat{{\bf{B}}_s}$) provides an \emph{asymptotically unbiased} ST-MUSIC signal subspace estimate up to at least $O\left( {{N^{ - 1}}} \right)$ terms. 

This claim is not evidently shared by earlier ST-MUSIC estimators \cite{DICLAUDIO13,DICLAUDIO14} that retain $O\left( {{N^{ - 1} }} \right)$ signal subspace components in $E\left\{ {{{\bf{A}}^H}{{{\bf{\hat E}}}_v}{\bf{\hat E}}_v^H{\bf{A}}} \right\}$, related to the source spectra. In fact, the different asymptotic bias at various frequencies is another source of DOA estimate instability at high SNR in wide-band applications without a regularized ${\bf{\hat \Gamma }}$. 

However, beside these finite sample effects, the asymptotic variances of the various ST-MUSIC subspaces are identical, as for the narrow-band MUSIC DOA estimator \cite{STOICA90a,MCCLOUD02,MESTRE08}. 

For the leakage components to be discarded, instead, ${{\bm{\Sigma }}_\varepsilon^2\left( {k,k} \right)}$ is generally close to one according to \eqref{eqn:ST-MUSIC}, so ${\mu _k}$ is comparable to the Capon PSD ${\left[ {{{\bf{B}}}{{\left( {:,k} \right)}^H}{{\bf{A}}^H}{{{\bf{\hat E}}}_s}{\bf{\Phi \Gamma \Phi \hat E}}_s^H{\bf{A}}{{\bf{B}}}\left( {:,k} \right)} \right]^{ - 1}}$, which is generally much greater than $c$ \cite{NADAKUDITI08}. However, in rare scenarios, leakage residuals might generate some $\mu_k$ of order $c$, interpreted as \emph{extremely weak} ghost sources, that have a vanishing impact for $N\rightarrow\infty$ and can be therefore harmlessly included in $\hat{\bf{B}}_s $. 

Finally, the $M - \kappa$ columns of $\bf B$, corresponding to the discarded $\mu_k \gg c$, are collected in the matrix $\hat{{\bf{B}}_\varepsilon}$.

\subsection{ST-MUSIC Rank Selection by AIC and BIC}
\label{sec:STMUSICRankSelectionbyAICandBIC}
Due to the high spectral variability of wide-band sources, the signal subspace rank $\kappa$ generally changes with frequency and should be estimated from data. The heuristic rules based on the singular value magnitude of ${\bf{A}}{\left( \nu  \right)^H}{\hat{\bf{E}}_s}$, used in earlier ST-MUSIC approaches \cite{DICLAUDIO13}, caused ambiguities.

On the contrary, the proposed ML formulation admits the use of Information Theoretic Criteria \cite{AKAIKE74,WAX85,BURNHAM04} for estimating $\kappa$. To this purpose, sample ${\mu _k}$s are ordered in a non decreasing manner\footnote{The quantity $\ln {\bm{\Sigma }}_s^2\left( {k,k} \right) + {c^{ - 1}}{{{\bm{\Sigma }}_\varepsilon^2\left( {k,k} \right)} \mathord{\left/
 {\vphantom {{{\bm{\Sigma }}_\varepsilon^2\left( {k,k} \right)} {{\bm{\Sigma }}_s^2\left( {k,k} \right)}}} \right.
 \kern-\nulldelimiterspace} {{\bm{\Sigma }}_s^2\left( {k,k} \right)}}$ increases monotonically with ${\mu _k}$ and ${\bm{\Sigma }}_\varepsilon^2\left( {k,k} \right)$.} and a set of nested models with reduced rank $K$ is compared for $0 \le K < M$. The log-likelihood of each model is written as
\begin{equation}
	\begin{array}{c}
{\varphi \left( K \right) } = K\left( {MP - \eta } \right)\ln \left( {\pi N_P^{ - 1}} \right) + \\
\left( {MP - \eta } \right)\ln \det \left[ {{\bm{\Sigma }}_s^2\left( {1:K,1:K} \right)} \right] + cK{\mathop{\rm trace}\nolimits} \left( {\bf{\Gamma }} \right) + \\
{N_P}{\mathop{\rm trace}\nolimits} \left[ {{\bm{\Sigma }}_v^2\left( {1:K,1:K} \right){\bm{\Sigma }}_s^{ - 2}\left( {1:K,1:K} \right)} \right] 
\end{array}
\label{eqn:likelihoodRR}
\end{equation}
where, by a continuity argument, ${\varphi \left( 0 \right) } = 0$. The estimates of $\kappa$ by AIC and BIC are readily \cite{AKAIKE74,WAX85,BURNHAM04} obtained as
\begin{equation}
	{\hat\kappa _{AIC}} = \mathop {\arg \min }\limits_{0 \le K \le M-1} \left\{ {2{\varphi \left( K \right) } + 2K\left( {2M - K + 1} \right)} \right\}
	\label{eqn:AICbinrank}
\end{equation}
\begin{equation}
	{\hat\kappa _{BIC}} = \mathop {\arg \min }\limits_{0 \le K \le M-1} \left\{ {2{\varphi \left( K \right) } + K\left( {2M - K + 1} \right) \ln \left( {MP - \eta } \right)} \right\} 
	\label{eqn:BICbinrank}
\end{equation}
since the number of free real parameters for the model order $K$ is $M^2$ for the preliminary whitening GSVD transformation (i.e., an unessential constant overhead), plus $K\left( {2M - K} \right)$ for the rank $K$ subspace ${{{\bf{\hat B}}}_s}$ \cite{GOLUB89}, plus $K$ for the generalized eigenvalues\footnote{The columns of ${\bf{B}}$ are orthonormal after the whitening and $\mu_k$, ${\bf{\Sigma }}_\varepsilon^2\left( {k,k} \right)$ and ${\bf{\Sigma }}_s^2\left( {k,k} \right)$ are all linked one-to-one.}. 

Either \eqref{eqn:AICbinrank} and \eqref{eqn:BICbinrank} implicitly define an environment dependent upper threshold $T_\mu > c$ for a valid $\mu_k$. Their detailed analysis is complicated by the statistical interaction with the prior estimation of $\eta$ and is left out of the scope of this paper. However, since $c \ll 1$ and $\mu_k \gg c$ for leakage components, \eqref{eqn:AICbinrank} and \eqref{eqn:BICbinrank} furnished reliable and almost indistinguishable detection results in simulation for both $\kappa$ and DOA estimation. This result further supports the findings and the basic assumptions made in the STCM analysis.

The estimates of $\kappa$ abruptly shrink at very low SNR, as in the sample SCM case \cite{NADAKUDITI08}. However, the eigenvectors discarded by \eqref{eqn:AICbinrank} and \eqref{eqn:BICbinrank} are certainly dominated by whitened leakage residuals, are numerically wobbly and must be excluded from ${{{\bf{\hat B}}}_s}$.
 
On the other hand, \eqref{eqn:likelihoodRR} involves a small ratio between the number of observations $MP-\eta$ and $\kappa < M$ and the risk for DOA estimation of including \emph{marginal} $\mu_k$s originated by false alarms is low. Therefore, the consistency claims of BIC for ${{\left( {MP - \eta } \right)} \mathord{\left/
 {\vphantom {{\left( {MP - \eta } \right)} {\kappa  \to \infty }}} \right.
 \kern-\nulldelimiterspace} {\kappa  \to \infty }}$ are weak and the slightly more permissive AIC \eqref{eqn:AICbinrank} may alleviate the $\kappa$ shrinkage issue at low SNR and was adopted in simulations.

\subsection{Optimal ST-MUSIC Subspace Weighting}
\label{sec:OptimalSTMUSICSubspaceWeighting}
Optimal DOA estimation in the WSF framework \cite{STOICA90a,VIBERG91} entails the calculus at each frequency of interest of the spatial \emph{perturbation covariance} ${{\bf{\Xi }}_\varepsilon } = {\rm{E}}\left\{ {{{{\bf{\hat B}}}_s}{\bf{\hat B}}_s^H} \right\} - {\rm{E}}\left\{ {{{{\bf{\hat B}}}_s}} \right\}{\rm{E}}\left\{ {{\bf{\hat B}}_s^H} \right\}$, where ${\rm{E}}\left\{ {{{{\bf{\hat B}}}_s}} \right\} = {{\bf{B}}_s}$ from \eqref{eqn:signalsubspaceeigenvalue} and of the optimal \emph{subspace weighting matrix} ${\bf{C}}_s $ of size $\kappa \times \kappa$, which whitens the perturbation of ${{\bf{B}}_s}$ onto $E\left\{ {{{{\bf{\hat B}}}_\varepsilon }} \right\} = {{\bf{B}}_\varepsilon }$. 

This task requires the $O\left( {{N^{ - {1 \mathord{\left/ {\vphantom {1 2}} \right. \kern-\nulldelimiterspace} 2}}}} \right)$ perturbative analysis of \eqref{eqn:signalsubspaceeigenvalue}, adapted and simplified from \cite{GOLUB89,PILLAI89,PAIGE84}. The eigenvector perturbation ${{\bf{\dot B}}_s}$ is herein decomposed for convenience as
\begin{displaymath}
	{{{\bf{\dot B}}}_s} = {\hat{\bf{B}}_s} - {{{\bf{B}}}_s} = {{{\bf{B}}}_s}{{\bf{Y}}_s} + {{{\bf{B}}}_\varepsilon }{{\bf{Y}}_\varepsilon }
\end{displaymath}
where ${{\bf{I}}_{\kappa}} + {{\bf{Y}}_s}$ describes a random rotation of ${\bf{B}}_s$, since all signal eigenvalues $\mu_k \rightarrow c$ asymptotically. 

Differentiating \eqref{eqn:signalsubspaceeigenvalue}, ${{\bf{Y}}_\varepsilon }$ asymptotically leads to
\begin{equation}
	 - \left( {{{{\bf{\dot \Pi }}}_\varepsilon } - c{{{\bf{\dot \Pi }}}_s}} \right){{{\bf{B}}}_s} \simeq {\rm{E}}\left\{ {{{\bf{\Pi }}_\varepsilon } - c{{\bf{\Pi }}_s}} \right\}{{{\bf{ B}}}_\varepsilon }{{\bf{Y}}_\varepsilon }
	\label{eqn:differentiation}
\end{equation}
where, after tedious calculus using \eqref{eqn:perturbedWSS}, \eqref{eqn:WLSweight}, \eqref{eqn:GSVD1}, \eqref{eqn:GSVD2}
\begin{equation}
	\begin{array}{c}
{{{\bf{\dot \Pi }}}_\varepsilon } \simeq {{\bf{\Pi }}_\varepsilon } - {\mathop{\rm E}\nolimits} \left\{ {{{\bf{\Pi }}_\varepsilon }} \right\} \simeq {{\bf{A}}^H}{{\bf{E}}_s}\left[ { - {{\bf{I}}_\eta } + 0.5 c{\bf{\Gamma }}} \right]{{\bf{\Gamma }}^{{1 \mathord{\left/
 {\vphantom {1 2}} \right.
 \kern-\nulldelimiterspace} 2}}}{\bf{G}}_0^H{\bf{E}}_v^H{\bf{A}}\\
 + {{\bf{A}}^H}{{\bf{E}}_v}{{\bf{G}}_0}{{\bf{\Gamma }}^{{1 \mathord{\left/
 {\vphantom {1 2}} \right.
 \kern-\nulldelimiterspace} 2}}}\left[ { - {{\bf{I}}_\eta } + 0.5 c{\bf{\Gamma }}} \right]{\bf{E}}_s^H{\bf{A}}
\end{array}
\label{eqn:deltaPieps}
\end{equation}
\begin{equation}
	\begin{array}{c}
{{{\bf{\dot \Pi }}}_s} \simeq {{\bf{\Pi }}_s} - {\mathop{\rm E}\nolimits} \left\{ {{{\bf{\Pi }}_s}} \right\} \simeq {{\bf{A}}^H}{{\bf{E}}_s}{{\bf{\Gamma }}^{{3 \mathord{\left/
 {\vphantom {3 2}} \right.
 \kern-\nulldelimiterspace} 2}}}{\bf{G}}_0^H{\bf{E}}_v^H{\bf{A}}\\
 + {{\bf{A}}^H}{{\bf{E}}_v}{{\bf{G}}_0}{{\bf{\Gamma }}^{{3 \mathord{\left/
 {\vphantom {3 2}} \right.
 \kern-\nulldelimiterspace} 2}}}{\bf{E}}_s^H{\bf{A}} \;.
\end{array}
\label{eqn:deltaPis}
\end{equation}

Inserting \eqref{eqn:GSVDas1}, \eqref{eqn:GSVDas2}, \eqref{eqn:deltaPieps} and \eqref{eqn:deltaPis} into \eqref{eqn:differentiation} and taking into account from \eqref{eqn:signalsubspaceeigenvalue} that ${\bf{E}}_v^H{\bf{A}}{{{\bf{ B}}}_s} = {\bf{0}}$, we get
\begin{equation}
	\begin{array}{*{20}{c}}
{{{\bf{A}}^H}{{\bf{E}}_v}{{\bf{G}}_0}{{\bf{\Gamma }}^{{1 \mathord{\left/
 {\vphantom {1 2}} \right.
 \kern-\nulldelimiterspace} 2}}}\left( {{{\bf{I}}_\eta } + 0.5c{\bf{\Gamma }}} \right){\bf{E}}_s^H{\bf{A}}{{{\bf{ B}}}_s} \simeq }\\
{\left[ {1 - N_P^{ - 1}{\rm{trace}}\left( {{\bf{\Gamma }} + c{\bf{\Phi }}{{\bf{\Gamma }}^2}{\bf{\Phi }}} \right)} \right]{{\bf{A}}^H}{{\bf{E}}_v}{\bf{E}}_v^H{\bf{A}}{{{\bf{ B}}}_\varepsilon }{{\bf{Y}}_\varepsilon }\;.}
\end{array}
\label{eqn:NSSperturbation}
\end{equation}

Taking expectations on both sides yields ${\mathop{\rm E}\nolimits} \left\{ {{{\bf{Y}}_\varepsilon }} \right\} = {\bf{0}}$, because ${\mathop{\rm E}\nolimits} \left\{ {{{\bf{G}}_0 }} \right\} = {\bf{0}}$, confirming the absence of asymptotic bias. In addition, right multiplying both sides by the unknown ${\bf{C }}_s$, neglecting terms independent of $\nu$, using \eqref{eqn:GSVDas2} and solving for ${{\bf B}}_\varepsilon {{\bf{Y}}_\varepsilon }{\bf{C }}_s$ asymptotically yields
\begin{equation}
\begin{array}{*{20}{c}}
{{{\bf{\Xi }}_\varepsilon } \propto {{{\bf{ B}}}_s}{\rm{E}}\left\{ {{{\bf{Y}}_\varepsilon }{{\bf{C}}_s}{\bf{C}}_s^H{\bf{Y}}_\varepsilon ^H} \right\}{{{\bf{ B}}}_s}^H = {{\left( {{{\bf{A}}^H}{{\bf{E}}_v}{\bf{E}}_v^H{\bf{A}}} \right)}^\dag } \times }\\
{N_P^{ - 1}{\rm{trace}}\left[ {{{\bf{C}}_s}{\bf{C}}_s^H{\rm{E}}\left\{ {{\bf{\Sigma }}_s^2\left( {1:\kappa ,1:\kappa } \right)} \right\}} \right]\;.}
\end{array}
\label{eqn:NSScov}
\end{equation}

Since from \eqref{eqn:signalsubspaceeigenvalue} ${\mathop{\rm E}\nolimits} \left\{ {{\bf{\Sigma }}_s^2\left( {1:{\kappa},1:{\kappa}} \right)} \right\} \simeq 
{\left( {1 + c} \right)^{ - 1}}{{\bf{I}}_{{\kappa}}} $, minimization of ${\rm{trace}}\left[ {{{\bf{C}}_s}{\bf{C}}_s^H{\rm{E}}\left\{ {{\bf{\Sigma }}_s^2\left( {1:\kappa ,1:\kappa } \right)} \right\}} \right]$, subject to $\det \left( {{{\bf{C}}_s}{\bf{C}}_s^H} \right) = 1$, asymptotically yields \cite{VIBERG91}
\begin{equation}
	{{{\bf{C }}_s}} = {{\bf{I}}_{{\kappa }}} 
	\label{eqn:Psis}
\end{equation}
or any other unitary matrix. From \eqref{eqn:signalsubspaceeigenvalue} a consistent estimate of ${\left( {{{\bf{A}}^H}{{\bf{E}}_v}{\bf{E}}_v^H{\bf{A}}} \right)^\dag }$ is ${\hat{\bf{B}}_\varepsilon }{\bf{\Sigma }}_\kappa ^{ - 2}\hat{\bf{B}}_\varepsilon ^H$, where 
\begin{displaymath}
{\bf{\Sigma }}_\kappa ^2 = \left( {1 + c} \right){\bf{\Sigma }}_\varepsilon ^2\left( {\kappa + 1:M,\kappa + 1:M} \right) - c{{\bf{I}}_{M - \kappa }} \;.
\end{displaymath}

Such an error covariance matrix \emph{confined} within the $\hat{\bf{B}}_\varepsilon$ subspace of dimension $M-\kappa$ can cause numerical troubles to WSF estimators especially during the coarse DOA initialization. So the sample ${{{\bf{\hat \Xi }}}_\varepsilon }$ can be completed by adding a rather arbitrary Hermitian matrix spanning the orthogonal complement ${{\bf{Q}}_\varepsilon }$ of ${\hat{\bf{B}}_\varepsilon }$. In particular, the choice 
\begin{equation}
	{{{\bf{\hat \Xi }}}_\varepsilon } = \kappa \left( {{\hat{\bf{B}}_\varepsilon }{\bf{\Sigma }}_\kappa ^{ - 2}\hat{\bf{B}}_\varepsilon ^H + {\zeta _\varepsilon}{{\bf{Q}}_\varepsilon }{\bf{Q}}_\varepsilon ^H} \right)
	\label{eqn:Xi}
\end{equation}
where ${\zeta _\varepsilon} = {\left( {M - \kappa } \right)^{ - 1}}{\mathop{\rm trace}\nolimits} \left( {{\hat{\bf{B}}_\varepsilon }{\bf{\Sigma }}_\kappa ^{ - 2}\hat{\bf{B}}_\varepsilon ^H} \right)$, effectively minimized the distance ${\left| {{{{\bf{\hat \Xi }}}_\varepsilon } - \kappa {\zeta _\varepsilon}{{\bf{I}}_M}} \right|_F}$ in most cases with ${\zeta _\varepsilon} \approx 1$. However, in the presence of strong leakage ${\hat {\bf{\Xi }}_\varepsilon }$ can still be ill-conditioned, hampering for instance the statistical efficiency of common focusing schemes \cite{DICLAUDIO05}.

\subsection{DOA estimation from the ML ST-MUSIC signal subspace}
\label{subsec:DOA estimation from the ML ST-MUSIC signal subspace}

Given the associate ${\hat {\bf{\Xi }}_\varepsilon }\left( \nu \right)$ and ${\bf{C}}_s\left( \nu \right)$, the sample ML ST-MUSIC subspace $\hat{\bf{B}}_s \left( \nu \right)$ at frequency $\nu$ can replace the SCM counterpart in any subspace based narrow-band DOA estimator. For instance, the optimal WSF estimator minimizes over the DOA parameter set ${\bf{\Theta }} = \left\{ {{{\bm{\theta }}_1}, \ldots ,{{\bm{\theta }}_D}} \right\}$ the functional
\begin{equation}
	{{\bf{\hat \Theta }}_{WSF}} = \mathop {\arg \min }\limits_{{\bf{\Theta }},{\bf{C}}} \left| {{{\widehat {\bf{\Xi }}}_\varepsilon }{{\left( \nu  \right)}^{ - \frac{1}{2}}}\left[ {{\bf{B}}\left( {{\bf{\Theta }},\nu } \right){\bf{C}} - {\hat{\bf{B}}_s}\left( \nu  \right){{\bf{C}}_s}\left( \nu  \right)} \right]} \right|_F^2
\label{eqn:ST-MUSIC_WSF}
\end{equation}
where ${\bf{B}}\left( {{\bf{\Theta }},\nu } \right)$ is the tentative steering matrix for ${{\bf{B}}_{11}}\left( \nu  \right)$ in \eqref{eqn:generalWSS}, ${\bf{C}}_s \left( \nu  \right) = \bf I_\kappa$ from \eqref{eqn:Psis} and $\bf{C}$ is an unknown full rank mixing matrix \cite{VIBERG91}. 

The extension of \eqref{eqn:ST-MUSIC_WSF} to multiple frequencies is straightforward \cite{CADZOW90}.  It allows consistent and asymptotically efficient DOA estimation in a Gaussian scenario at $\nu$, thanks to the ML formulation of the ST-MUSIC subspace and of the WSF minimum DOA variance property \cite{VIBERG91}. Consistency holds for any fourth-order bounded scenario in the absence of other model mismatches \cite{SWINDLEHURST92,SWINDLEHURST93,DICLAUDIO01}. 

The estimator \eqref{eqn:ST-MUSIC_WSF} analyzes \emph{spectral slices} of the STCM \cite{CAPON69} with a \emph{very narrow effective bandwidth} around $\nu$, comparable to the width of a null of \eqref{eqn:loglikelihood0}, virtually extrapolating the STCM and eliminating the effects of steering vector changes with frequency, beside spectral leakage. 

Following this argument, working on $P$ equi-spaced frequencies may not optimally exploit the source information by ML ST-MUSIC, since some strong narrow-band components might be missed. In particular, \eqref{eqn:columnspaceofEs} indicates that an optimal ST-MUSIC would require the coherent analysis of \emph{at least} $P+L_d-1$ optimally selected points of the full $z$ plane (i.e., by using a \emph{complex} $\nu$) \cite{WAX84,MOHAN08}. 

In addition, since the consistency of signal subspace fitting does not require the \emph{exact} specification of ${{\bf{\Xi }}_\varepsilon }\left( \nu \right)$ and ${{{\bf{C }}_s}}\left( \nu \right)$ \cite{VIBERG91,DICLAUDIO01}, the $O\left( {{N^{ - 1/2}}} \right)$ estimates \eqref{eqn:Psis} and \eqref{eqn:Xi} are adequate. However, obtaining robustness to steering vector mismatches at $\nu$ may require different choices of \eqref{eqn:Psis} and \eqref{eqn:Xi} \cite{SWINDLEHURST92,SWINDLEHURST93,DICLAUDIO01}. 

In contrast, with reference to \eqref{eqn:generalWSS}, the energy from adjacent frequencies in the span of ${{\bf{B}}_{11}}\left( \nu  \right)$ (i.e., the component ${{\bf{B}}_{11}}\left( \nu  \right){\bf{B}}_{11}^\dag \left( \nu  \right){{\bf{B}}_{12}}\left( \nu  \right){{\bf{C}}_{F2}}\left( \nu  \right)$) is retained by the SCM and canceled by the ML ST-MUSIC, which may give to Fourier methods a detection advantage at low SNR with a coarse frequency binning. 

The overall statistical efficiency of the ML ST-MUSIC subspace, which, as the SCM, is a \emph{reduced statistic} w.r.t. the STCM, remains an open question. However, working in an ideal narrowband scenario (i.e., after posing ${\bf{A}} = {{\bf{I}}_M}$), the weightings of the ML ST-MUSIC and of the optimal narrowband subspaces \cite{VIBERG91} are proportional. 

The STCM allows more flexible strategies in non-stationary environments. For instance, a STCM can be built from many disjoint blocks of few STSs each to locate intermittent, elusive sources without sacrificing the spectral resolution and the effective SNR as in the binning approach \cite{HARRIS78}. 

\section{Computational analysis}
\label{sec:ComputationalAnalysis}

The ML ST-MUSIC computing cost is largely dominated by the building and the EVD of the STCM. In particular, the STCM computation requires a bulk of $M^2 NP$ multiply and accumulate operations using the apparatus developed for fast AR identification \cite{KAYMARPLE81}. The dominant EVD cost of the sample STCM is about $6M^3P^3$ complex flops. Viable alternatives include fast Lanczos algorithms \cite{GOLUB89}.

This effort must be compared to the about $6M^3 P_w+NM^2$ flops required by the SCM approach for the same tasks. To give an idea, in simulations using MATLAB®, the full WAVES using ML ST-MUSIC ran in about $1.0$ s, while the SCM based WAVES in about $25$ ms for $N=6400$, $M=8$, $P=P_w=64$ and $33$ analyzed frequencies on a PC machine featuring an Intel Core i7-6700K processor running at $4.2$ GHz and $32$ GB RAM. However, in most real-time settings, the marginal cost of STCM processing is comparable with the overhead of coarse DOA initialization techniques using high performance wide-band beamformers \cite{KROLIK89,DICLAUDIO03,BUCRIS12}, orthogonal matching pursuit for WSF \cite{CADZOW90,VIBERG91}, or refined focusing schemes \cite{LEE94,DICLAUDIO01,DICLAUDIO05,BUCRIS12}.
 

\section{Computer simulations}
\label{sec:results}

The performance of ST-MUSIC was assessed by Monte-Carlo time domain simulations\footnote{It is worth noting that inconsistency effects cannot be observed if wide-band array outputs are simulated in the frequency domain by a set of independent narrow-band signals.} of a ULA with $M=8$ omni-directional sensors, equi-spaced by $0.5$ wavelengths at the center frequency $\nu_0$. The sensor bandwidth ranged from $0.6 \nu_0$ to $1.4 \nu_0$. The background Gaussian noise was temporally and spatially white (${{\bf{R}}_{vv}} = {{\bf{I}}_{MP}}$) and the source SNR was referred to each array element. 

For each SNR value, $1000$ independent Monte Carlo trials were run, collecting $N=6400$ sensor snapshots sampled at $T = {\left( {0.8{\nu_0}} \right)^{ - 1}}$ s and using $P=64$ ($N_P = 6337$) for the STCM and an un-windowed DFT of $P_w = P$ points over $N_s = 100$ non overlapping, consecutive data segments for SCM based approaches\footnote{Setting $P_w = P$ allowed the use the same focusing matrices in Sects. \ref{subsec:Widebandfocusing} and \ref{sec:WideBandARSources}. Due to the previous discussion, these choices may not be optimal for each estimator and scenario.} \cite{HUNG88,STOICA90a,DICLAUDIO01}. 

Four sources emitting various kinds of signals were placed at the azimuth angles $8^\circ$, $13^\circ$, $33^\circ$ and $37^\circ$, referred to the array broadside. Performance comparison employed state of the art WSF type DOA estimators, namely MODE \cite{STOICA90a} and WAVES \cite{DICLAUDIO01} followed by MODE, in turn fed by sets of signal subspaces drawn by ML ST-MUSIC and by SCM, optimally weighted for finite sample errors according to \cite{VIBERG91,DICLAUDIO01}, \eqref{eqn:Psis} and \eqref{eqn:Xi}. Since DOA variance and bias have different dominant causes (finite sample errors and model mismatches, respectively \cite{DICLAUDIO05}), they were separately analyzed. 

\subsection{Single frequency analysis}
\label{subsec:Single frequency analysis}
This experiment tested the \emph{consistency} and the basic accuracy of the subspace estimates. Sources radiated equi-powered and uncorrelated wide-band Gaussian noise, filtered in the band $\left( {0.725{\nu_0},1.275{\nu_0}} \right)$ (see Fig. \ref{fig:Fig2}). MODE (i.e., a root version of \eqref{eqn:ST-MUSIC_WSF}) was applied in turn to the ML ST-MUSIC subspace estimate at $\nu_0$, to the corresponding SCM estimate at bin $0$ and to the reference subspace drawn from a \emph{classical} narrow-band SCM \cite{SCHMIDT86} built with $N_s = 100$ independent snapshots, whose Cramer Rao bound (CRB) for DOA was also calculated. For the STCM, $\eta = 200$ was selected from the median of the BIC estimates at the highest SNR of $40$ dB. The ST-MUSIC subspace rank $\kappa$ was selected online by AIC \eqref{eqn:AICbinrank}. The signal subspace rank of the sample SCM and the number of sources searched by MODE were instead fixed at four.
\begin{figure}
	\centering
		\includegraphics[width=3.1in]{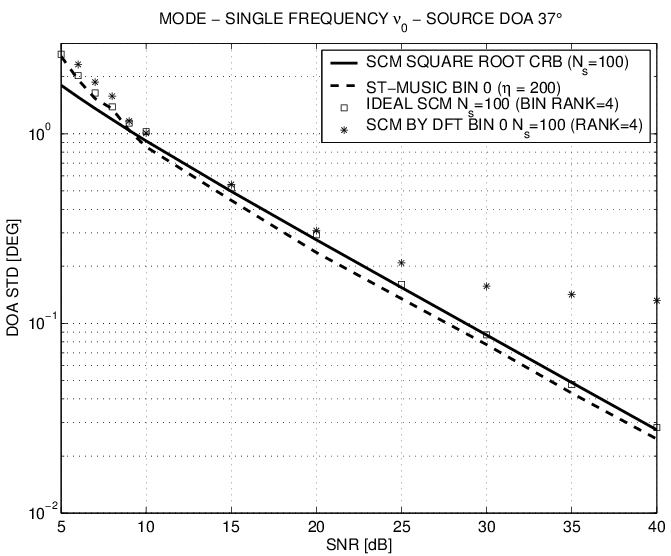}
	\caption{Plot of the DOA sample standard deviation versus the SNR for the source impinging from $37^\circ$, observed at the single frequency $\nu_0$, obtained by MODE \cite{STOICA90a} applied to different signal subspace estimates, compared with the narrow-band SCM square root CRB for $N_s=100$.}
	\label{fig:Fig3}
\end{figure}
\begin{figure}
	\centering
		\includegraphics[width=3.1in]{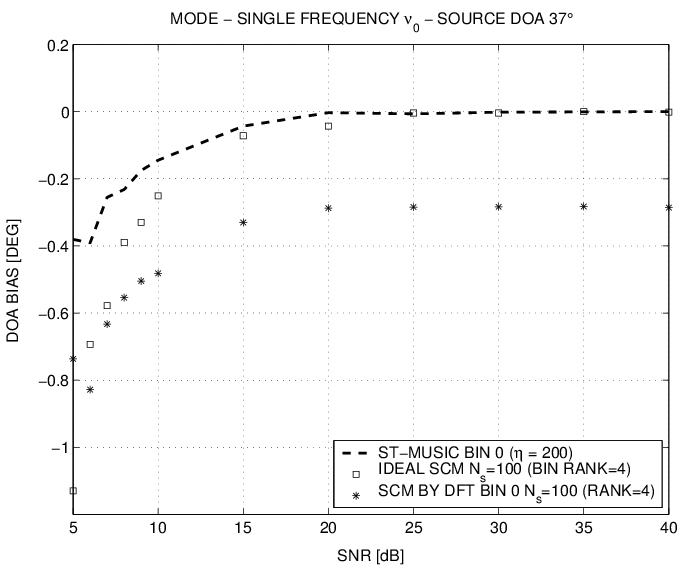}
	\caption{Plot of the DOA sample bias versus the SNR for the source impinging from $37^\circ$, observed at the single frequency $\nu_0$, obtained by MODE applied to different signal subspace estimates.}
	\label{fig:Fig4}
\end{figure}

DOA standard deviation and bias are displayed in Figs. \ref{fig:Fig3} and \ref{fig:Fig4} versus the SNR for the source impinging from $37^\circ$, the most critical one for CRB and steering vector change with frequency. For all algorithms, the low SNR threshold was at around $5$ dB. The DFT-generated subspace was clearly non-consistent. The ML ST-MUSIC extracted more information than the classical SCM subspace for $N_s=100$, obtaining lower DOA variance, coupled with negligible and similar bias. These results support the consistency and the superior accuracy potential of STCM and ML ST-MUSIC and can be directly extended to multi-frequency WSF estimators \cite{CADZOW90}.
\subsection{Wide band focusing}
\label{subsec:Widebandfocusing}
The test ran in the same scenario, but focusing the center frequencies of the $33$ bins in the band $\left( {0.8{\nu_0},1.2{\nu_0}} \right)$ onto $\nu_0$ using WAVES \cite{DICLAUDIO01}, followed by MODE. In this case, DOA consistency was basically hampered by the focusing bias \cite{DICLAUDIO05,YOON06} and the test emphasized the quality of the information combination across frequencies and the robustness to model errors. Unitary focusing matrices were used, assuming that preliminary beamforming estimated two DOA clusters centered at $10.50^\circ$ and $35^\circ$ with focusing sectors as in \cite{HUNG88,DICLAUDIO01}.\cite{HUNG88}. However, the number of focusing angles was increased to twelve ($6.7^\circ$, $8.6^\circ$, $10.50^\circ$, $12.3^\circ$, $14.1^\circ$, $31^\circ$, $32.3^\circ$, $33.6^\circ$, $35^\circ$, $36.3^\circ$, $37.6^\circ$ and $39^\circ$) for better accuracy of array interpolation. In fact, the unitary Procrustes DOA interpolation problem \cite{HUNG88,GOLUB89,DORON92} over the sector ensemble $S$ yields a focusing matrix equal to the orthogonal polar factor of $\int\limits_{{\bm{\theta }} \in S} {{\bf{b}}\left( {{\nu _0},{\bm{\theta }}} \right){\bf{b}}{{\left( {\nu ,{\bf{\theta }}} \right)}^H}d{\bm{\theta }}}$. The original angle choice of \cite{HUNG88} did not adequately cover the effective sector rank and approximate this integral \cite{DORON94A}. 

The focusing accuracy was still unsatisfactory for our purposes at high SNR and forced us to set the WAVES rank to four. As an additional measure, a cascaded, non-unitary sector interpolation matrix \cite{FRIEDLANDER93}, unique for all bins, remapped the principal singular vector of each set formed by the $33$ focused steering vectors at the same DOA onto the corresponding ULA steering vector at $\nu_0$\cite{DICLAUDIO05SP}. This heuristic correction substantially reduced the bias and still satisfied Fisher optimality criteria \cite{HUNG88}. 

The CRB and the SCM based WAVES bound under the \emph{perfect subspace mapping} (PSM) assumption \cite{DORON92,DICLAUDIO01} were computed for $N_s = 100$ independent snapshots drawn from each DFT bin. Moreover, beside the cases of fixed selection for $\eta = 200$ and the SCM rank, two further experiments were included, where these quantities were all estimated online by AIC \cite{WAX85}.
\begin{figure}
	\centering
		\includegraphics[width=3.1in]{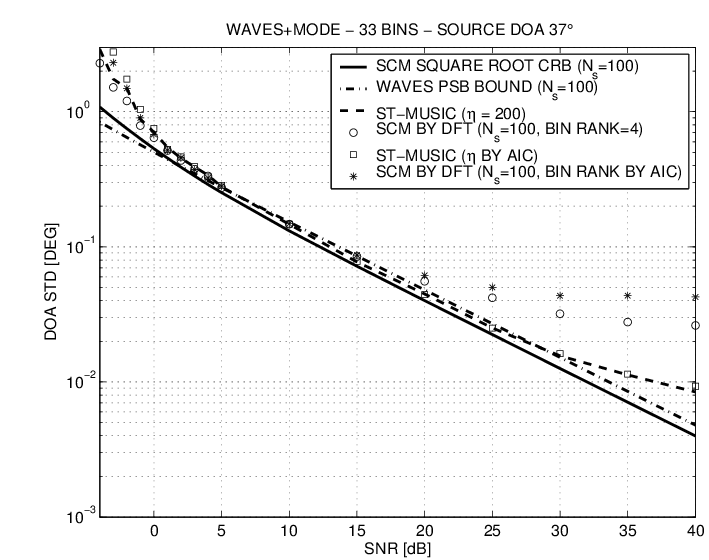}
	\caption{Plot of the DOA sample standard deviation versus the SNR for the source impinging from $37^\circ$, obtained by WAVES \protect \cite{DICLAUDIO01}, focused on $33$ bin frequencies, followed by MODE applied to different signal subspace estimates, and compared with the narrow-band SCM square root CRB and the WAVES PSM bound computed for $N_s=100$.}
	\label{fig:Fig5}
\end{figure}
\begin{figure}
	\centering
		\includegraphics[width=3.1in]{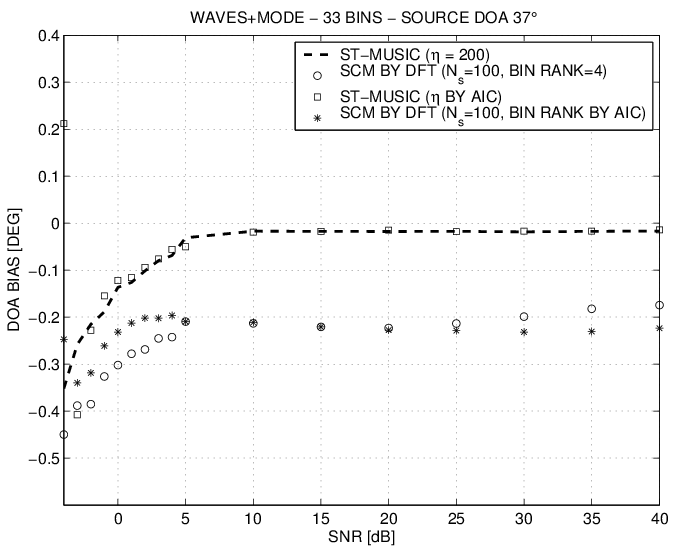}
	\caption{Plot of the DOA sample bias versus the SNR for the source impinging from $37^\circ$ for the various subspace estimates.}
	\label{fig:Fig6}
\end{figure}

Sample results were displayed in Figs. \ref{fig:Fig5} and \ref{fig:Fig6} for the source at $37^\circ$. The standard deviation plot of Fig. \ref{fig:Fig5} revealed a slight advantage of SCM over ML ST-MUSIC subspaces at low SNR, with a threshold for reliable detection slightly higher for AIC driven rank estimates. The difference appears essentially due to a slightly higher (not removed) outlier rate of ML ST-MUSIC at extremely low SNR, because the AIC detected rank $\kappa$ shrunk at several frequencies in some trials. Instead, the overall insensitivity of ML ST-MUSIC to the AIC selection of $\eta$ was remarkable, as expected.

The SCM based WAVES exhibited the usual breakdown at high SNR, further complicated by the ghost source detection when using AIC, and was also heavily biased at every SNR. The DOA variance of the WAVES with ML ST-MUSIC subspaces appeared slightly degraded at extreme SNRs w.r.t. the single frequency case. The most likely causes are the focusing errors and the Fisher information loss \cite{HUNG88} due to the suboptimal averaging of ${\hat {\bf{\Xi }}_\varepsilon }\left( \nu \right)$ across the pass-band, performed by the chosen \emph{fixed} focusing scheme \cite{DICLAUDIO01}.

\subsection{Wide Band Coherent Sources}
\label{sec:WideBandCoherentSources}

\begin{figure}
	\centering
		\includegraphics[width=3.1in]{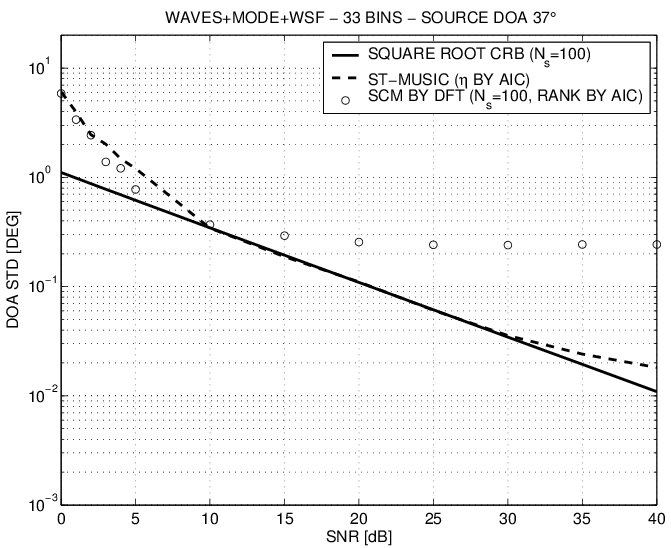}
	\caption{Plot of the DOA sample standard deviation versus the SNR for the coherent source impinging from $37^\circ$, obtained by WAVES, focused on $33$ bin frequencies, followed by MODE+WSF, applied to different signal subspace estimates and compared with the narrow-band SCM square root CRB computed for $N_s=100$.}
	\label{fig:Fig9}
\end{figure}

\begin{figure}
	\centering
		\includegraphics[width=3.1in]{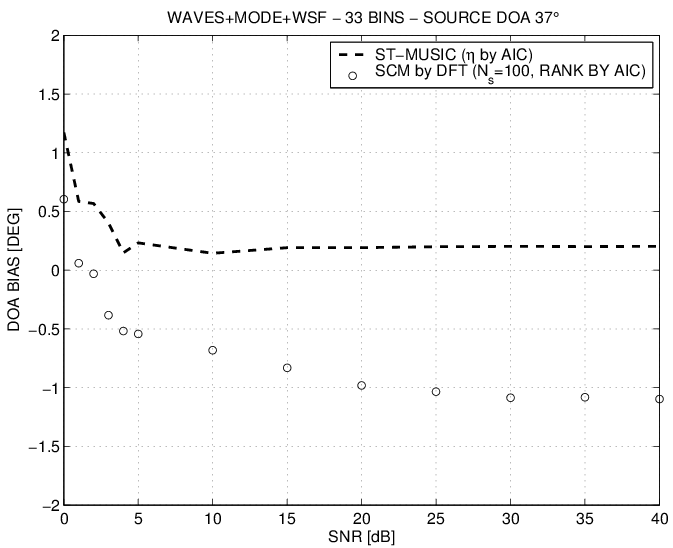}
	\caption{Plot of the DOA sample bias versus the SNR for the coherent source impinging from $37^\circ$ for the various subspace estimates.}
	\label{fig:Fig10}
\end{figure}

The test was repeated with similar results after replacing the sources at $13^\circ$ and $37^\circ$ with delayed replicas of those at $8^\circ$ and $33^\circ$ with delays $0.45T$ and $0.35T$ respectively \cite{DICLAUDIO13}. In these trials, there was not any obvious choice for $\eta$ and the SCM rank, that were selected by AIC. MODE had local convergence issues and its DOA estimates were refined by WSF Newton iterations. Results are shown in Figs. \ref{fig:Fig9} and \ref{fig:Fig10}. The PSB was not shown since it is ambiguous for coherent arrivals, depending on the focusing strategy.

\subsection{Wide Band Autoregressive Sources}
\label{sec:WideBandARSources}
The previous test was repeated after replacing the white sources with independent $16$-th order autoregressive (AR) sources, whose models were drawn from live recordings at an airport, assuming $\nu_0 = 2 \pi \times 10^3$ rad/s. The source correlation length ranged from $30T$ to $50T$. The SNR was referred to the driving Gaussian noise variance, equal for all sources. The STCM rank estimated by BIC at high SNR was $\eta = 199$ in this environment.

\begin{figure}
	\centering
		\includegraphics[width=3.1in]{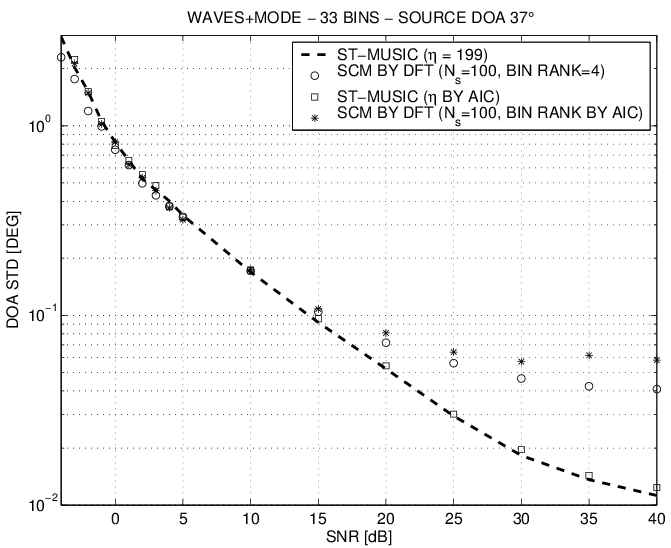}
	\caption{Plot of the DOA sample standard deviation versus the SNR for the AR source impinging from $37^\circ$, obtained by WAVES, focused on $33$ bin frequencies, followed by MODE, applied to different signal subspace estimates.}
	\label{fig:Fig7}
\end{figure}

\begin{figure}
	\centering
		\includegraphics[width=3.1in]{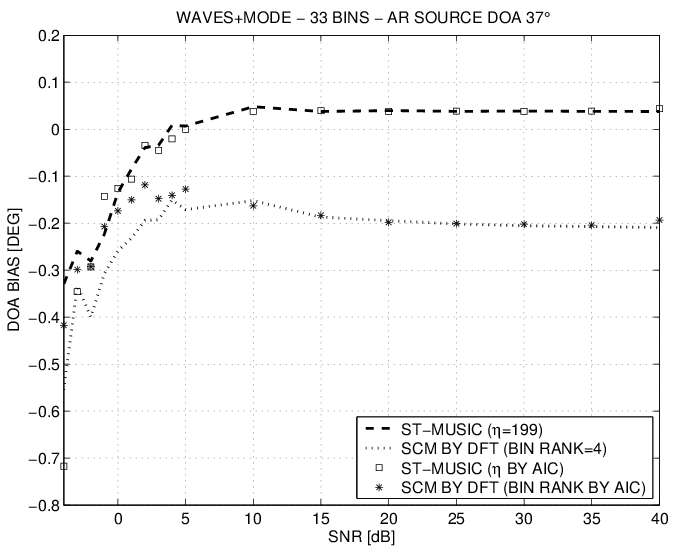}
	\caption{Plot of the DOA sample bias versus the SNR for the AR source impinging from $37^\circ$, for the different signal subspace estimates.}
	\label{fig:Fig8}
\end{figure}

Sample results were displayed in Figs. \ref{fig:Fig7} and \ref{fig:Fig8} for the source at $37^\circ$ and confirmed previous findings.


\section{Conclusion}
\label{sec:conc}

A consistent signal subspace estimator at arbitrary frequencies (ML ST-MUSIC) has been developed by exploiting the finite rank representation of wide-band sources in the signal subspace of the STCM through a ML inverse fitting, based on the MUSIC paradigm. AIC and BIC estimators of the number of sources active at each frequency of interest have been naturally derived in this framework. The ML ST-MUSIC approach radically circumvents the frequency spread issues of classical binning and is theoretically supported by the asymptotic nulling of the spectral leakage, identified as the source of the statistical inconsistency. Optimal weighting for finite sample errors allows the use of ML ST-MUSIC subspaces in any narrow- or wide-band subspace fitting DOA estimator. Simulations support the consistency, the superior accuracy (evident at high SNR) and the robustness to mismatches (evident in automatic model selection) of DOA estimation derived from ML ST-MUSIC subspaces w.r.t. the SCM based counterparts.

Finally, the ML ST-MUSIC derivation raises the delicate question if binning approaches are theoretically justified even in narrow-band scenarios. In fact, since it is always $L_d > 1$ in physical arrays, a STCM with low order $P$ should be used in order to guarantee consistency. 

\ifCLASSOPTIONcaptionsoff
 \newpage
\fi



\bibliographystyle{IEEEtran}
\bibliography{IEEEabrv,imageprocessing,arrayprocessing}

%
%
%

%

\ifCLASSOPTIONdraftcls
\newpage
\else{\ifCLASSOPTIONtwoside{



\vfill

}
\else
\newpage
\fi}
\fi


\end{document}